\documentclass[useAMS,usenatbib,useasmath]{mn2e}
\usepackage{epsfig}
\usepackage{longtable}
\usepackage{times}
\usepackage{amsmath}
\usepackage[usenames,dvipsnames]{color}

\newcommand{\Rmax}{\hbox{$R_{\hbox{\scriptsize \rm max}}$}}
\newcommand{\Rmin}{\hbox{$R_{\hbox{\scriptsize \rm min}}$}}
\newcommand{\Rcore}{\hbox{$R_{\hbox{\scriptsize \rm core}}$}}

\newcommand{\Lsun}{\hbox{$L_\odot$}}
\newcommand{\Msun}{\hbox{$M_\odot$}}

\newcommand{\Teff}{\hbox{$T_{\hbox{\small eff}}$}}

\newcommand{\NV}{\hbox{N$_{\hbox{\scriptsize V}}$}}
\newcommand{\ND}{\hbox{N$_{\hbox{\scriptsize D}}$}}

\newcommand{\chir}{\hbox{$\chi_{\hbox{\small R}}$}}
\newcommand{\chip}{\hbox{$\chi_{\hbox{\small P}}$}}

\newcommand{\tage}{\hbox{$t_{\hbox{\small age}}$}}

\newcommand{\kms}{\hbox{km$\,$s$^{-1}$}}

\newcommand{\tsub}[1]{\hbox{$t_{\scriptsize \rm  #1}$}}

\newcommand{\deriv}[2]{ {\partial #1 \over \partial #2} }

\def\lesssim{\mathrel{\hbox{\rlap{\hbox{\lower4pt\hbox{$\sim$}}}\hbox{$<$}}}}
\def\gtrsim{\mathrel{\hbox{\rlap{\hbox{\lower4pt\hbox{$\sim$}}}\hbox{$>$}}}}
\def\gray{$\gamma$-ray}
\def\grays{$\gamma$-rays}
\def\isoni{$^{56}{\rm Ni}$}
\def\isoco{$^{56}{\rm Co}$}
\def\ha{H$\alpha$}
\def\one{{\,\sc i}}

\def\fig{Fig.}
\def\figs{Figs.}
\def\eq{Eq.}
\def\eqs{Eqs.}
\def\rte{radiative-transfer equation}
\def\rtes{radiative-transfer equations}

\title[Time Dependent Radiative Transfer Calculations for Supernovae]
{Time Dependent Radiative Transfer Calculations for Supernovae}

\author[ D. John Hillier,  Luc Dessart]
{D. John Hillier$^1$\thanks{E-mail: hillier@pitt.edu}, Luc Dessart $^{2,3}$ \\
$^1$ Department of Physics and Astronomy \& Pittsburgh Particle physics, Astrophysics, and Cosmology Center (PITT PACC), University of Pittsburgh, Pittsburgh, PA 15260, USA \\
$^2$: Laboratoire d'Astrophysique de Marseille, Universit\'e Aix-Marseille \& CNRS,
UMR7326, 38 rue Fr\'ed\'eric Joliot-Curie, 13388 Marseille, France \\
$^3$ Theoretical Astrophysics, MC 350-17, California Institute of Technology, Pasadena, CA 91125, USA}

\begin{document}

\date{Accepted . Received }

\pagerange{\pageref{firstpage}--\pageref{lastpage}} \pubyear{2011}

\maketitle

\label{firstpage}

\begin{abstract}
In previous papers we discussed results from fully time-dependent radiative transfer models for core-collapse supernova (SN) ejecta, 
including  the Type II-peculiar SN 1987A,  the more ``generic'' SN II-Plateau, and more recently Type IIb/Ib/Ic SNe. Here we describe 
the modifications to our radiative modeling code, {\sc cmfgen}, which allowed those studies to be undertaken. 
The changes allow for time-dependent radiative transfer of SN ejecta in homologous expansion. 
In the modeling we treat the entire SN ejecta, from the innermost layer that does not fall back on the compact remnant out to the progenitor surface layers. From our non-LTE time-dependent line-blanketed synthetic spectra, we compute the bolometric and multi-band light curves: light curves and spectra are thus calculated simultaneously using the same physical processes and numerics.
These upgrades, in conjunction with our previous modifications which allow the solution of the time dependent rate equations, 
will improve the modeling of SN spectra and light curves, and hence facilitate new insights into SN ejecta properties, the SN progenitors and the explosion mechanism(s). {\sc cmfgen} can now be applied to the modeling of all SN types.  
\end{abstract}


\section{Introduction}

Supernova (SN) spectra potentially contain a great deal of information about the progenitor star, about the explosion dynamics, and nucleosynthesis yields. Extracting this information,  particularly at early times, is difficult due to their low densities and high velocities at their effective photosphere. Because of the low densities, radiative processes tend to  dominate over collisional processes and hence local-thermodynamic equilibrium (LTE) cannot be assumed. Instead we must solve the statistical  equilibrium equations, and this requires vast amounts of atomic data, much of which has only become available over the last decade. The high velocities blend spectral features together,  often making line identifications difficult, and this hinders spectral analysis since accurate spectral modeling is needed in order to interpret weak, but important, diagnostics.

An inherent feature of SNe is that their spectra evolve with time, on a timescale comparable to, or shorter than, the age of the SN. Thus a crucial question is whether this time dependence needs to be taken into account when modeling SN spectra.\footnote{When modeling light curves there are no controversies --- time dependency is essential for modeling the light curves of SNe. An excellent discussion of some of the issues regarding time-dependence is provided by \cite{PE00_Ia_anal}. A related issue is the necessity of including all terms of  order $v/c$ in the transfer equation; \cite{MM84_RH} provide an in-depth discussion of the importance of various terms in the transfer equation.} Can reliable results from spectral fitting be obtained by modeling only the photospheric layers, in the spirit of radiative-transfer studies on stellar atmospheres? That is, can we ignore time-dependent terms, and can we impose a lower boundary condition and adjust this boundary condition, and photospheric abundances,  until our model spectrum matches what is observed? This approach has been adopted by many workers in the field, following the first papers by \cite{BDN85_SN_spec}, \cite{Lucy87_ASN87A} and \cite{MLD93_MC_Ia}.


For example, \cite{DH05_epm} used the technique to explore the expanding photosphere method (EPM) for finding distances to Type II SNe.  With these investigations they  were able to investigate the physics influencing  the EPM technique, and hence understand why it was difficult to accurately estimate the correction factors using photometry  alone. \cite{DH06_SN1999em} applied the EPM technique to SN 1999em, and showed that the distance obtained agreed within 10\% of the cepheid distance.  Later, \cite{DBB08_SN2005cs} successfully applied the EPM method to determine the distance to SN 2005cs and SN 2006bp. Our modeling of  these three Type II-P SNe did, however, reveal a problem that is also seen in modeling of  other hydrogen-rich SNe  --- the inability to predict H$\alpha$ during the  recombination epoch (after approximately day 20 in  Type II-P SNe). At this and later times the model H$\alpha$ line is consistently narrower and weaker than observed.

The \ha\ problem is also seen for SN 1987A; however in this case problems in fitting H$\alpha$ occur as early as day 3 \cite[e.g.,][]{SAR90_87A} due to the more rapid cooling of the ejecta from a more compact progenitor.  \cite{SAR90_87A} postulated that clumping might be the source of the H$\alpha$ discrepancy. \cite{EK89_87A} have also modeled SN 1987A (up to day 10) and did not find any H$\alpha$ discrepancy. Two effects might contribute to this difference. First, \cite{EK89_87A} treated many metals in LTE which could affect the H ionization balance. Second, as shown by \cite{EK89_87A}, the H$\alpha$ emission strength is sensitive to the density exponent $n$ defined by $n=-d\ln \rho/d\ln r$. Low n (i.e., $n=7$) give much stronger H$\alpha$ emission than higher $n$ (i.e., $n=20$).  \cite{SAR90_87A} used a power law with a density exponent of 10, while \cite{EK89_87A} used a density exponent of 9. In the study of \cite{DH06_SN1999em} it was found that while lowering the density exponent down from 10 to 8, and even to 6, led to an increase in the H$\alpha$ strength it also produced a very severe mismatch of the model with observations of the Ca\,{\sc ii} lines.

A related problem is that \cite{SAR90_87A} were unable to fit the He\,{\sc i} emission in SN 1987A -- models predicted no He\,{\sc i}  $\lambda$5876 although it is easily seen in the observations (up to day 4). Simple model changes did not improve the predictions, and hence they postulated an extra source of UV photons. The models of  \cite{EK89_87A} also underestimated He\,{\sc i} $\lambda$\,5876. They suggested a combination of effects as the likely cause --- high helium abundance, X-rays produced in an interaction zone, density effects, and time-dependence effects.

\cite{MBB01_mixing} found that models for 1987A, computed using the  code {\sc phoenix} \citep{HB99_phoenix,SHB99_PHOENIX}, also  predicted H$\alpha$  emission much weaker than that observed. To explain these discrepancies they invoked mixing 
of \isoni\ into the SN's outer layers --- mixing at velocities ($v > 5000\,$\kms) larger than predicted in current hydrodynamic modeling \citep{KPS06_mix,JWH09_mix,HJM10_3D_SN_sims}\footnote{The recent 3D simulations of \cite{HJM10_3D_SN_sims} do produce nickel mixing out to 4500\,\kms, but this is insufficient.  Further, in order to explain H$\alpha$, which is the most significant spectral feature in Type II SN spectra, isolated bullets will not suffice -- there has to be global mixing into the hydrogen photosphere.}. The H\,{\sc i} and He\,{\sc i} lines would be excited by non thermal electrons produced by the degradation of gamma-rays created by the decay of radioactive nickel. \cite{BNB03_nick_mix} also used  mixing of $^{56}$Ni into the SN's outer layers to explain the H\,{\sc i} and He\,{\sc i} lines in early-time spectra of SN 1993W (Type II-P).

The H\,{\sc i} and He\,{\sc i} discrepancies are important --- either there is a fundamental problem with our current spectral modeling of SNe, or there is crucial SN physics missing from current hydrodynamic models which leads, for example, to an underprediction of the amount and extent of \isoni\ mixing. In order to distinguish these possibilities it is important that we relax assumptions, such as steady state, in our SN   models. Indeed, \cite{UC05_time_dep} showed that it was important to include time dependence in the rate equations for SN 1987A. When such terms are included, it is possible  to explain the large strength of the Balmer lines without invoking non-thermal excitation resulting from the mixing of radioactive nickel from below the hydrogen-rich envelope.  This conclusion has been confirmed by \cite{DH08_time} who showed that inclusion of the time dependent terms allowed the H$\alpha$ line strength to be better matched in 
SN 1999em, particularly at late times ($> $20 days).


The necessity of including time-dependent terms is unfortunate. With steady-state models we can readily adjust the luminosity, abundances and the density structure to tailor match the observations. Instead we have an initial value problem requiring us to begin the calculations at a sufficiently early time such that the adopted initial model allows the full time evolution to be modeled. On the other hand, this necessity will allow us to better test both hydrodynamical models of the explosions, and models for the progenitor evolution.

More recently  \cite{DH10_time}, extended their work to allow for time dependence of the radiation field. This removes issues with the inner boundary condition, and controversies on which terms can be omitted from the \rte\ \citep{BHM96_cmf, PE00_Ia_anal}. With their fully time-dependent approach, they were able to successfully model  the early evolution of SN 1987A, from $\lesssim$1\,d up to 20\,d. Overall, predicted spectra were in good agreement with observation, and showed the expected trend from a very high effective temperature at early times ($\Teff > 30,000$K at 0.3\,d) to a relatively constant effective temperature at later times ($\Teff \sim 6000$\,K at 5\,d).  The latter is a consequence of the hydrogen ionization front which sets the location of the SN photosphere.  One deficiency in the models was an excess flux in the $U$ band. This excess flux could arise from a deficiency in the model calculations (e.g., insufficient line blanketing) or from deficiencies in the initial model (e.g., wrong sized progenitor). This work, with time dependent radiative transfer and time dependent rate equations, showed that standard models for 1987A, without \isoni\ mixing beyond 4000\,\kms  could explain both the H\,{\sc i} and He\,{\sc i} lines in  SN~1987A. The result is in agreement with explosion models which require mixing out to $\sim 3000$ to $4000\,$\kms\ in order to explain the observed light curve of SN 1987A \citep[e.g.,][]{SN90_SN1987A_lc, BLB00_SN87A_lc}, the appearance of hard X-rays \cite[e.g.,][]{KSN89_87A_xrays}, and later nebular diagnostics \citep[e.g.,][]{HMT09_SN2005bl}.

Time dependent radiation transfer has also been included in {\sc phoenix} \citep{JHB09_Dt}, as has the solution of the time-dependent rate equations \citep{SBH10_dSSEdt}. A distinction between {\sc phoenix} and {\sc cmfgen} is that {\sc phoenix} works primarily with the intensity, while {\sc cmfgen} solves for the radiation field using moments. The latter method has two advantages: It explicitly allows for electron scattering, and in the time-dependent approach a factor of $N_{\scriptsize \rm angle} (\sim$\ND) less memory is required to store the time-dependent radiation field. A disadvantage is the generalization to multi-dimensions, and stability issues.

In our work, we solve the \rte, or its moments,  numerically. An alternative technique is to use a Monte-Carlo approach. A major proponent of the Monte-Carlo technique is Lucy, who has devoted considerable effort to its development \citep[e.g.,][]{Luc99_RE_MC, Luc99_SN_MC, Luc02_MC, Lu03_MC, L05_MC}. Techniques have been developed that allow the temperature structure of the envelope to be solved for \citep{Luc99_RE_MC, BW01_RE_MC}. Most, if not all, the Monte-Carlo codes use the Sobolev approximation for the line transfer; this should be an excellent approximation in SN atmospheres although  \cite{BHM96_cmf} argue that the approximation is suspect in SNe because of the larger number of lines whose intrinsic profiles overlap. While it is possible to undertake full non-LTE calculations with Monte-Carlo codes, such calculations are very time consuming. Thus it is usual to make approximations to determine the ionization structure [e.g., LTE or a modified nebular approximation;  \cite{AL85_winds}] and/or the line source function (e.g., LTE).

Early applications of Monte-Carlo methods to SNe are discussed by \cite{ML93_MC} and were applied to the modeling of SN Ia spectra by \cite{MLD93_MC_Ia}. Since that time the code has undergone numerous extensions \citep[e.g.,][]{M00_MC} and has been applied to interpreting spectra of many SNe \cite[e.g.,][]{MDP06_hyper,MDH09_SnIb}. The code is not time dependent, and generally uses a  gray photosphere approximation for the inner boundary condition. Deviations in the level populations and ionization structure are taken into account using a nebular approximation \citep{MDH09_SnIb}. These Monte-Carlo codes are neither LTE, nor fully non-LTE. Since line photons are allowed to scatter, non-LTE is partially treated even when LTE populations are assumed.

A more recent example of a Monte-Carlo code is {\sc sedona} \citep{KTN06_SN_MC}.{\sc sedona} calculates emergent spectra, as a function of time, assuming LTE level populations. Line transfer for the ``strongest lines" (up to 0.5 million) are treated using the Sobolev approximation \cite{Sobolev60,Cat70_sob} while all remaining lines are treated using an expansion opacity formalism \citep{KLC77_eop,EP93_SN_ALI}. In {\sc sedona} photons can be either absorbed by a line, scatter, or fluoresce (cause an emission in another transition but which has the same upper level). In the study of the width-luminosity relation for Type Ia SNe redistribution processes were taken into account using a two-level atom approach with a constant redistribution probability of 0.8 \citep{KW07_Ia_width}  whereas in another study it was set to 1 \citep[i.e., pure absorption;][]{WKB07_Ia_lc}\footnote{In many papers it is referred to as the thermalization parameter, and can be thought of as the fraction of line photons that are ``absorbed'' on a given line interaction \citep[e.g.][]{HWS92_nova}. This probability not only allows for true absorption, but also, for example, re-emission of a photon from the upper level in another bound-bound transition.}. One advantage of the Monte-Carlo technique is the relative ease with which time-dependence is treated \citep{L05_MC}.
More recent work using Monte-Carlo techniques has been undertaken by  \citet{maurer_etal_11} and \citet{jerkstrand_etal_11}  who solve the non-LTE time-dependent rate equations, and solve for the radiation field during the nebular phase. The approach allows for a more accurate treatment of the emergent flux for quantitative analyses especially since, in the nebular phase, the coupling between the gas and the radiation is relatively weak. Monte-Carlo codes tend to be inefficient in 1D but have the advantage that they can be readily extended to 2D and 3D.  Given the complexities of SN modeling it is essential that we have alternative approaches whose results can be compared. 



In the present paper we outline the assumptions underlying our time-dependent radiative transfer calculations, and describe our solution technique. In our work we have assumed that the SN flow can be adequately described by a homologous expansion (i.e., $v \propto r$). There are several reasons for doing this. First, SNe approach a homologous state at late times. For example, Type Ia SNe rapidly expand to a radius of $\lesssim$\,10$^{14}$\,cm from an object the size of the Earth ($\lesssim$\,10$^9$\,cm) in one day. After such a rapid expansion, the assumption of a homologous expansion is excellent. This is also true for Type Ib, Ic SNe which expand, from an object similar in size to the Sun, by over a factor of 100 in 1 day.  For Type II SN the assumption becomes a good approximation after a few days for a blue-supergiant (BSG) progenitor and somewhat later ($\sim 20$\,d) for a red-supergiant (RSG) progenitor. Several factors contribute to the departure at early times. First  the progenitor radius is large (particularly for a RSG), and it takes time for the expanding ejecta to ``forget" its initial radius. Second, the energy density of the radiation field at early times is significant compared with the kinetic energy, and this influences the hydrodynamics \citep{falk_arnett_77,DLW10a}.
Third, in Type II SNe, the reverse shock can set up a non-monotonic velocity field between the core and He/H interface. 
Finally, material can fall back towards the core, which is particularly important for high mass progenitors.


The second reason for neglecting the departure from a homologous expansion is that in this complex, but still exploratory work, we do not wish to be concerned with dynamics.  A third reason is that the radiation transfer equation is simpler and there are computational advantages in solving this equation compared with the full transfer equation.  In particular, the third moment of the radiation field does not appear in the moment equations.

This paper is organized as follows: In \S\ref{time_effects} we briefly characterize how time dependence can affect the modeling of SNe. The \rte, its moments and associated boundary conditions in space and time, are discussed in \S\ref{Sec_rad_trans}.  The time dependent rate and radiative equilibrium equations are discussed in \S\ref{Sec_stat} while the deposition of radioactive energy is discussed in \S\ref{Sec_rad_energy}. A brief description of the numerical method used to handle the Lagrangian derivative is provided in \S\ref{Sec_num_sol}. As in other {\sc cmfgen} modeling, we use a partial linearization method to facilitate the simultaneous solution of the \rte, and the rate and radiative equilibrium equations (\S\ref{Sec_linearization}).
The use of super-levels, which reduces the size of the non-LTE model atoms,  and problems introduced by the use of super-levels are discussed in \S\ref{Sec_SLs}. Model convergence is discussed in \S\ref{Sec_convergence} while code testing is  discussed in \S\ref{Sec_testing}. The computation of the observed spectrum, which is done in both the comoving and observer's frames, is  discussed in \S\ref{Sec_comp_obs_spec}. In \S\ref{Sec_gray} we discuss the time dependent gray transfer equations, which can be used to provide an initial estimate of the temperature structure. From these equations we derive a global energy constraint  (\S\ref{Sec_glob_en}). Finally, in \S\ref{Sec_future}, we discuss additional work that is needed to improve 1D modeling of SNe.

\section{Time Dependent Effects in SN\lowercase{e}}
\label{time_effects}

Time dependence enters into the \rte\ because the speed of light is finite.
To characterize its importance, we can consider three separate cases:

\begin{enumerate}

\item
Consider an atmosphere with a scale height $\Delta R$. In the optically thin limit, the characteristic light travel time is simply  $\Delta R/c$. 
For a SN, the characteristic flow time is $R/v$ and hence the ratio of the light travel time to the flow time is 
$$  {\Delta R \over R } {v \over c} $$
which for $v=10,000\,\kms$ and   $\Delta R / R=1/10$ (appropriate for $\rho = \rho_o (r_o/r)^{10}$) we have $\tsub{light}/\tsub{flow} \sim 1/300$. Because this is much less than 1, it is generally assumed that the time dependency of the \rte\ can be ignored when modeling the photosphere. This will generally be reasonable even when the scale height of the photosphere  approaches the local radius.\\

\item
 In the optically thick regime the effective light travel time will increase since photons cannot freely  stream --- instead they perform a spatial random walk, mediated by absorption and scattering processes. In this random walk, photons will travel  a distance $\Delta R$  from their origin on a timescale $\tau \Delta R /c$, and hence $$\tsub{light}/\tsub{flow} = { \tau \Delta R \over R } {v \over c}\,\,.$$ That is, the random walk of the photons through the medium increases the light travel time by a factor of $\tau$. Thus, as soon as the optical depths become large (e.g., $\tau> 100$) the random-walk time will approach, or exceed, the flow timescale.  In this case, the time dependence terms are crucial, and will control the temperature structure of the SN ejecta. The preceding statement  applies to all SN ejecta and their light curves, i.e., optical depth effects systematically require a time-dependent treatment. However, the effect is most pronounced, and lasts longer, in those SNe (i.e., Type IIP)  with large ejecta masses.


The random-walk time scale is equivalent to the diffusion time-scale, but the use of the word diffusion in this context is somewhat of a misnomer, since ``classical'' diffusion normally refers to the case where we have a temperature gradient. However, in core-collapse SNe the temperature gradient is imposed by the shock, and may be positive, negative, or zero. At large optical depths the photons are trapped by the optically thick ejecta. If the SN were not expanding, some of these photons would eventually ``diffuse'' to the surface, and would impose a temperature gradient on the ejecta. However, in the ``optically thick'' regions of Type II SN ejecta the expansion time is shorter than the diffusion time, and the temperature is set by the influence of expansion (adiabatic cooling) on the initial temperature structure, and photon escape from depth is primarily determined by the contraction of the photosphere to lower velocities. Initially the light curve reflects  the cooling induced by adiabatic expansion; later the light curve is controlled by the escape  of thermal energy as the recombination front moves into the SN ejecta \citep[e.g.,][]{EWW94_IIp,Arn96_book}.

The relative importance of the different energy sources will depend on the mass of $^{56}$Ni created, and the mass and the initial radius of the ejecta. For example, Type Ia SNe differ significantly from Type II SNe. In Type Ia SNe the explosion occurs in such a small object (approximately the size of the Earth) that the initial explosion is never seen\footnote{\cite{hoeflich_schaefer_09} have studied whether the explosion could be seen in X-rays and gamma-rays. Their studies suggest that the Burst and Transient Source Experiment (BATSE) should have seen about $13 \pm 4$ nearby SNe Ia while none were detected. They suggest that absorption by the accretion disk can account for this discrepancy. The optical light curve, arising from the shock breakout, should be detectable, with an optical/UV luminosity of order $10^6$ to $10^7$\,\Lsun \citep{PCW10_Ia_SBO,RLW11_Ia_shock}.}
---- rather the entire visible light curve of Type Ia SNe  is dictated by the re-heating of the SN ejecta by \isoni/\isoco\ decay. In the {\it unmixed} Type IIb/Ib/Ic models of  \cite{DHL11_Ibc} the re-brightening caused by the diffusion heat wave  arising from the decay of $^{56}$Ni was not seen until about 10 days --- prior to this the light curve exhibited a plateau. Models without $^{56}$Ni showed a similar plateau,  but subsequently faded. The light curve of SN 1987A, because it stems from the explosion of a BSG star, is initially controlled by adiabatic cooling, but then, as in Type Ia SNe, becomes dominated by the energy released from nuclear decay 
\citep[e.g.,][]{SN90_SN1987A_lc,BLB00_SN87A_lc}.

\item
A third case to consider is where the photospheric properties are changing rapidly. Even if $\tsub{light}/\tsub{flow} <1$, and light travel time effects are unimportant for the transport of radiation in the photosphere, light travel time effects can still influence observed spectra. This occurs because photons arising from an impact parameter $p=0$ (i.e., the observer's line of sight striking the SN center of mass) will arrive at the observer at a time $R_{\rm \scriptsize phot}/c$ earlier than photons arising from rays from an impact parameter with  $p=R_{\rm \scriptsize phot}$ (i.e., the observer's line of sight tangent to the photosphere).  An example of this effect can occur near shock breakout where the photospheric conditions are changing rapidly. A RSG has a radius of $\sim 10^{14}$ cm, and hence the travel time  across the radius of the star is 1h which is comparable to the variability timescale, and hence light-travel times effects would significantly effect the early light curve \cite[e.g.,][]{GDB08_GALEX,NS10_elc}. While the above discussion is primarily related to early photospheric phases, it may also be relevant at late times. At late times, the evolution of the properties of the SN ejecta and  light curve are controlled by the deposition of radioactive energy, and the relevant time scale to compare the light travel time with is the smaller of the decay time and the
flow time. For the photospheric phase, the dominant nuclear energy sources are the decays of $^{56}$Ni and $^{56}$Co, and these
have decay times of the same order as the flow time.

\end{enumerate}

In the following, we will be primarily concerned with effect (ii). Case (i) is difficult to handle 
because of the very short time scales. In our studies the effects are likely to be small, even allowing for the
extension of the SN ejecta. While we do treat time dependence, the large time steps (2\% to 10\% of the current time) means
that the (small) influence of freely traveling light is smoothed in time -- in  a 10\% time step the light can 
travel the whole of the grid. To accurately treat such an effect, if important, we would need to consider time steps
comparable to the light travel time across grid points. For our studies it is the flow time that is crucial,
since it is the primary factor determining the time scale on which the ejecta properties and radiation field
change with time. In the limiting case of a very thin photosphere (i.e., the atmosphere
can be treated in the plane-parallel approximation), case (iii) is more a bookkeeping exercise ---
the observed spectrum is simply the weighted sum of spectra from a series of steady-state
plane-parallel atmospheres computed at different time steps. That is, 
\begin{equation}
F_\nu(t_o + d/c) ={2\pi \over d^2} \int_0^{\Rmax}  p \\
  I^{+}_\nu(t_o - \Delta z/c , \Rmax, p) dp
\end{equation}
\noindent
with 
\begin{equation}
\Delta z=  \left(\sqrt{\Rmax^2-p^2}\right)
\end{equation}
\noindent
where $t_o$ is the time the observed light is emitted for an impact parameter $p=\Rmax$, $d$ is the distance to the SN, and for simplicity
we have ignored the expansion of the photosphere during the time interval $\Delta t=\Rmax/c$.

In our simulations \Rmax\ is the same for all frequencies, and is chosen so that at the outer boundary the SN ejecta 
is optically thin. While the latter condition cannot always be met (e.g., in the continuum of the ground state of a dominant ion
or for a strong resonance transition) extensive tests with W-R and O star models, and a lesser number of tests with SN models, show
that this does not affect the observed spectra in the important (i.e., the regions containing the flux) spectral regions (see \S~\ref{Sec_bound_cond} for further discussion). At early epochs, the outer boundary typically lies at a velocity of 0.1 to 0.2 $v/c$.

\section{Radiative Transfer}
\label{Sec_rad_trans}

The transfer equation in the comoving-frame, to first order in $v/c$, and assuming
a homologous expansion (so that $\partial v / \partial r = v/r$)  is
\begin{eqnarray}
{1 \over c}{\partial I_\nu \over \partial t} + {\mu c + v\over c}{\partial I_\nu \over \partial r}
&+& {(1-\mu^2) \over r} {\partial I_\nu \over \partial \mu} \nonumber \\
&-& {v \nu \over r c} {\partial I_\nu \over \partial \nu}
+ {3 v  \over r c} I_\nu
= \eta_\nu - \chi_\nu I_\nu
\label{eq_i_eqn}
\end{eqnarray}

\noindent
\citep{MM84_RH} where $I_\nu = I(t, r,\mu,\nu)$,  $\chi_\nu= \chi(t, r,\nu)$ is the opacity, $\eta_\nu=\eta(t, r,\nu)$ is the emissivity, and $I$, $\chi$, $\eta$, $\mu$, and $\nu$ are measured in the comoving-frame.
$\eta_\nu$ and $\chi_\nu$ are assumed to be isotropic in the comoving frame, and we have dropped the subscript $o$ (which denotes comoving frame) on $I$,  $\chi$, $\eta$, $\mu$, $\nu$  for simplicity.
The moments of the radiation field are defined in terms of the specific intensity   by

\begin{equation}
[J,H,K]= \int^{1}_{-1} [1,\mu,\mu^2] I(t,r,\mu,\nu) d\mu \,.
\end{equation}

\noindent
Integrating over the transfer equation, and rearranging terms, we obtain the zeroth and first moment of the \rte:

\begin{equation}
  {1 \over cr^3}  {D(r^3 J{_\nu})  \over Dt} + {1 \over r^2} {\partial (r^2 H_\nu)  \over \partial r}
  - {\nu v \over rc} { \partial J_\nu \over \partial \nu } = \eta_\nu - \chi_\nu J_\nu
 \label{eq_zero_mom}
 \end{equation}
\noindent
and

\begin{equation}
  {1 \over cr^3}  {D(r^3 H_\nu)  \over Dt} + {1 \over r^2} { \partial(r^2 K_\nu)  \over \partial r}
  + {K_\nu - J_\nu \over r} - {\nu v \over rc}{ \partial H_\nu \over \partial \nu } = - \chi_\nu H_\nu \,\,.
\label{eq_first_mom}
 \end{equation}

\noindent
where $D/Dt$ is the Lagrangian derivative which for spherical geometry is

\begin{equation}
\begin{split}
{D \over Dt} = \deriv{}{t} + v\deriv{}{r}\,\,.
\end{split}
\end{equation}


\noindent
An alternative formulation of  equations \ref{eq_zero_mom} and \ref{eq_first_mom} is
\begin{equation}
  {1 \over cr^4}  {D(r^4 J{_\nu})  \over Dt} + {1 \over r^2} {\partial (r^2 H_\nu)  \over \partial r}
  - {v \over rc} { \partial \nu J_\nu \over \partial \nu } = \eta_\nu - \chi_\nu J_\nu
 \label{eq_zero_mom_alt}
 \end{equation}
\noindent
and
\begin{equation}
  {1 \over cr^4}  {D(r^4 H_\nu)  \over Dt} + {1 \over r^2} { \partial(r^2 K_\nu)  \over \partial r}
  + {K_\nu - J_\nu \over r} - {v \over rc}{ \partial \nu H_\nu \over \partial \nu } = - \chi_\nu H_\nu \,\,.
\label{eq_first_mom_alt}
 \end{equation}

\noindent
These equations more directly show the connection to the gray problem (the frequency derivatives integrate to zero)
(see \S\ref{Sec_gray}) and the importance of adiabatic cooling due to expansion in the optically thick regime. That is $J \propto 1/r^4$ and hence $T \propto 1/r$.  We utilized \eqs~\ref{eq_zero_mom}--\ref{eq_first_mom} in our formulation, primarily because, in the absence of the D/Dt terms,  they are identical to those we use to solve the transfer equation in stellar winds (at least if $v \propto r$).

\subsection{The Eddington factors, \hbox{$f_\nu$}}
\label{Sec_edd_f}

The two moment equations discussed in \S\ref{Sec_rad_trans} (\eqs~\ref{eq_zero_mom} \& \ref{eq_first_mom}) contain 3 unknowns ($J_\nu$, $H_ \nu$, and $K_\nu$), and are thus not closed. To close these equations we introduce the closure relation $f_\nu=K_\nu/J_\nu$ \citep[see, e.g.,][]{Mih78_book} where $f_\nu$  is obtained from a formal solution of the transfer equation (Section~\ref{Sec_form_sol}). Since the formal solution depends on $S_\nu$, which in turn depends on $J_\nu$, we need to iterate. As is well known, convergence of this iteration procedure is rapid.

The Eddington factor, $f_\nu$, which provides a measure of the isotropy of the radiation field is a ratio of moments. As such they are relatively insensitive to the radiative transfer, and we therefore compute the Eddington factors using the fully relativistic, but time independent, moment equations.  By solving the time-dependent moment equations we correctly allow for the trapping of radiation in the optically thick SN envelope. Thus the temperature structure is correct and we compute the correct level populations. Near the surface we are neglecting light-travel time effects, but this will only be important in the event that it effects the angular distribution of the radiation, and in particular the Eddington factor $f_\nu$.

\subsection{Boundary Conditions}
\label{Sec_bound_cond}

\subsubsection{Outer boundary}

For the solution of the moment equations we assume
\begin{equation}
H_\nu=h_\nu J_\nu
\end{equation}

\noindent
at the outer boundary where $h$ is computed using the formal solution. This assumption is used in \eq~\ref{eq_first_mom} which is differenced in the usual way. The use of this boundary condition can sometimes generate instabilities in the solution, particularly at longer wavelengths (in the infrared). One manifestation of this instability is that a numerical/differencing error introduced into the solution may remain constant (or even grow). Since J declines with increasing wavelength, the percentage error in J increases. The instability is monitored using the formal solution, and can be reduced/eliminated by choosing a finer resolution at the outer boundary such that $\Delta \tau_1 << \Delta \tau_2$ where indices, starting at 1 at Rmax, are incremented inwards until ND at the innermost grid point..

In the formal (i.e., ray) solution we have two choices at the outer boundary. In the simplest approach we
adopt $I_\nu^-=0$ at the outer boundary. Depending on the size of the model grid, this choice can create model difficulties because $\tau_\nu>1$ at some frequencies implying that $I_\nu^- \sim S_\nu$. Using $I_\nu^-=0$ we obtain $J_\nu-S_\nu = S_\nu/2$ rather than $J_\nu-S_\nu << S_\nu$ for these frequencies, and this causes rapid changes in level populations at the outer boundary. In order to treat the boundary region correctly it would be necessary to resolve the outer boundary using many additional grid points. In practice this is unwarranted --- the outer boundary condition (provided it is reasonable) does not affect the solution, and the sharp truncation at the outer boundary is an artifact of the model construction.

As an alternative, we can set  $I_\nu^- =  0 $ at an extended outer boundary, and solve the transfer equation in this extended region to obtain $I^-$ at the true outer boundary. In SN models, the extended region typically extends a factor of 1.5 larger in radius, and opacities and emissivities in this region are obtained by extrapolation. In order to limit the velocity to be less than $c$ (the SN1987A models discussed by \cite{DH10_time} had $v(\Rmax)/c=0.205$ and $(R_{\hbox{\small ext}}/c=0.29$) we use a beta-law extrapolation of $v$ [$v(r)=(1-\Rcore/r)^\beta$] , rather than a pure homologous expansion. The technique of extrapolating the radius produces a much smoother behavior in the level populations at the outer boundary. It has been successfully used previously for the modeling of Wolf-Rayet stars, LBVs, and O stars \citep[e.g.,][]{Hil87_A, HM98_blank, HM99_WC}.

\subsubsection{Inner boundary}

For core-collapse SNe, the inner boundary represents the innermost ejecta layer that sits at the junction between the ejecta and fallback material (deemed to fall onto the compact remnant).  Depending on the model, and in particular the binding energy of the progenitor helium core \citep{DLW10b}, the inner ejecta velocity can vary between $\lesssim$\,100\,\kms for standard explosions while for energetic explosions, or explosions of low mass progenitors,  the velocities can be an order of magnitude larger. For Type Ia SNe, for which no remnant is left behind, the inner core velocities are on the order of  $\gtrsim1000\,\kms$ \citep{khokhlov_etal_93}, and still little mass travels at such low velocities compared to what obtains in Type II SN ejecta.

 For simplicity we adopt
\begin{equation}
H_\nu=0
\end{equation}

\noindent
(in actuality we tend to adopt a small non-zero value for $H_\nu$, but such that $H_\nu << J_\nu$). For models with an optically thick core, 
this assumption will produce a smooth run of level populations and temperature with depth near the inner boundary. Our assumption of 
$H_\nu=0$ ignores the possible influence of magnetar radiation etc. In reality (and assuming we have the correct SN density structure) the flux at the inner boundary is determined via symmetry arguments. The inward intensity at the inner boundary is the outward intensity at the inner boundary redshifted in frequency space and from an earlier time step. Thus
\begin{align}
I(t, r_{\rm min}, \mu_o, \nu_o,) = \notag \\
&({\nu_o / \nu_1})^3 I(t-2 \mu r_{\rm min}/c, r_{\rm min}, -\mu_o, \nu_1)
\end{align}
\noindent
with $\nu_1=\nu_o(1+\mu\beta)/(1-\mu \beta)$, and where we have ignored the distinction between $\mu$ and $\mu_o$ since $v << c$.

Thus  $H_\nu$ has a complicated frequency variation and would need to be determined in the iteration procedure.  However, when integrated over frequency,  $H \sim \beta J$ and thus from the gray perspective the assumption of $H=0$ is accurate. 

In practice it is reasonably easy to take the frequency shift into account. In the comoving-frame technique we iterate from blue to red, and thus the intensity information at the high frequency is always available. In the formal solution we can either use $H_\nu=0$ (and thus $I^+_\nu=I^-_\nu$) or we specify $I^+_\nu$ taking into account the frequency shift. Doing the later is of importance when the core is optically thin, and we are computing the observed spectrum. When computing an observed spectrum, and with an optically thin envelope,  the frequency shift is  potentially important for obtaining an accurate spectrum.

We have also implemented the non-zero $H_\nu$ case into the moment solutions. The straight forward approach of using $H_\nu=h_\nu J_\nu$ at all frequencies introduced  some numerical instabilities into the solution, and it was found that it was better to use this approach only in frequency regions where the core was optically thin.  We also find some convergence difficulties at the inner boundary as we transition from optically thick to optically thin, and these are still being investigated.  The cause of the difficulty is most likely related to resolution issues and the large variation in optical depth with frequency.  When $H_\nu$ is no longer small,  we effectively have a ``photosphere'' at the inner boundary, and in this ``photospheric'' region populations will change rapidly.

Accounting for the time difference in the inner boundary condition is more problematic (since the intensities at earlier time steps would be needed) but its effect is likely to be small since the relevant time scale is very small ($\Delta t \le 2\Rmin/c=2\tage\,(v(\Rmin)/c)$ where \tage\ is the time
since the SN exploded). Theoretically, it could  be taken into account using $I\_nu$ computed at previous time steps. 

Our adopted boundary condition is physically realistic, and is a much better approach than applying a boundary condition at some intermediate depth  below the photosphere. It also allows a better handling of the spectral evolution at late times when some spectral regions start becoming transparent while others remain thick.

\subsubsection{Time}
\label{sect_time}

The radiative transfer problem represented by \eqs~\ref{eq_i_eqn}, \ref{eq_zero_mom}, \& \ref{eq_first_mom} is an initial value problem in time.
Thus, to solve these equations, we need to fully specify $I_\nu(t_{init},r,\mu)$ and the corresponding moments. The computation of $I_\nu(t_{init},r,\mu)$ can only be done in conjunction with a full hydrodynamical simulation. 
This is beyond the scope of the present work. We therefore proceed as follows:
\begin{enumerate}
\item We adopt the temperature, density, and abundance structure from a full hydrodynamical model. In the past, Stan Woosley has supplied  us with such models. As these models, which are for RSG progenitors, are not fully resolved in the outer layers, we stitch a power law onto the hydrodynamical models at some suitably chosen velocity (e.g., 10,000\, to 30,000 \kms). As our choice may affect spectra, particularly at early times, we now obtain/compute models that more accurately describe the outer layers. The issue is less  of a problem for Type  Ia, Ib, and Ic SNe after one day, because the outermost layers of these models have been greatly diluted by the huge expansion the ejecta has undergone.

\item We solve for the radiation fields using a fully relativistic transport solver ignoring
time dependent terms ($\partial J_\nu/\partial t, \partial H_\nu /\partial t$). In performing this calculation we allow for non-LTE,
and we iterate to obtain the populations simultaneously with the radiation
field (i.e., we undertake a normal non-time dependent {\sc cmfgen} calculation). The temperature is held fixed at the value obtained from the hydrodynamical model.
\end{enumerate}
At depth this procedure is more than adequate for Type II SN, since $J_\nu$ is close to $B_\nu$.
However closer to the photosphere, effects may be seen since $J_\nu$ departs from $B_\nu$,
and the diffusion time can be significant relative to the age of the SN. The long term
effects of this inconsistency are probably small  but need to be examined. We can examine errors introduced by the starting solution by comparing {\sc cmfgen} results at some future time with the hydrodynamic solution at that time, and by comparing models at the same epoch but which begin with different starting solutions.

A second issue is  that the temperature structure from the hydrodynamical structure will not be exact --- typically the temperature has been determined assuming LTE and a gray-opacity approximation. Only on the second time step will the temperature structure of the outer layers adjust itself to be more fully consistent with our more accurate treatment of the radiation field,  and the non-LTE opacities and emissivities. Since the LTE approximation is valid for the inner layers, inconsistencies in the outer  temperature structure will only affect the earlier model spectra --- with time these outer layers become optically thin and no longer contribute to the observed spectrum. A comparison of the different temperature structures for Type Ib/Ic SN computed using different assumptions is shown by \cite{DHL11_Ibc}.

For the time step, we typically adopt $\Delta t=0.1t$. Tests show this is adequate over most of the time evolution. Exceptions will occur at shock breakout (not currently modeled) and at the end of the plateau phase in Type II SN as the photosphere passes from the hydrogen rich envelope into the He core. As the plateau stage ends, the luminosity can change by a factor of a few on a time scale of ~10\,days. As this is roughly 0.1 times the age of the SN, a smaller step size needs to be used. In the Type II-P core collapse  models of \citet{DH11_SNII} time steps of 2 to 5 days did a much better job of resolving the end of the plateau phase \cite[see Fig.~10 in][]{DH11_SNII}. At present the time step is set by hand, with 10\% being the default.

\subsection{Formal Solution}
\label{Sec_form_sol}

In order to solve for the moments of the radiation field we need to compute the Eddington factors, which are used to close the moment equations. These are computed using the  fully relativistic, but time independent, transfer equation. The fully relativistic transfer equation is
\begin{align}
{\gamma ( 1+ \beta\mu) \over c} {\partial I \over \partial t}& + \gamma(\mu +\beta)
{\partial I \over \partial r} \notag \\
& + \gamma(1 -\mu^2) \left[ { 1+\beta\mu \over r} - \Lambda \right] {\partial I \over \partial \mu} \notag \\
& -  \gamma \nu  \left[ {\beta (1-\mu^2) \over r} + \mu \Lambda \right]  {\partial I \over \partial \nu} \notag \\
& +  3 \gamma \left[ {\beta (1-\mu^2) \over r} + \mu \Lambda \right] I =
\eta -\chi I
\label{eq_full_rel_i}
\end{align}

\noindent
where $\beta=v/c$,
\begin{equation}
\gamma=1/\sqrt{1-\beta^2}\,\,\,,
\end{equation}
\begin{equation}
\Lambda= {\gamma^2 (1+\beta\mu) \over c}{\partial \beta \over \partial t} + {\gamma^2 (\mu+\beta)}{\partial \beta \over \partial r}
\end{equation}
and it is explicitly assumed that $I=I(t,r,\mu,\nu)$ with $I$, $\mu$ and $\nu$ measured in the comoving frame \citep{Mih80_rel_flow}.

To solve the transfer equation we will assume that we can ignore all time-dependent terms. This is reasonable, since we are not using this equation to compute the temperature structure of the ejecta. Rather, we will use its solution to compute the Eddington factors for the solution of the moment equations. As discussed in \S\ref{time_effects} and \S\ref{Sec_edd_f}, errors introduced by this approximation are likely to be small. For the computation of the observed spectrum, we are ignoring the light travel time for freely flowing photons, but the errors introduced by this assumption will generally be small, except near shock breakout (Section~\ref{time_effects}). The technique we use to solve \eq~\ref{eq_full_rel_i} closely follows that of   \cite{Mih80_rel_flow}.

Using the method of characteristics, the transfer equation can be reduced from a partial differential equation in three dependent variables $(r, \nu,  \mu)$ to a series of equations in two dependent variables $(s, \nu)$.  From \eq~\ref{eq_full_rel_i}  it follows that the characteristic equations are:
\begin{equation}
{dr \over ds} = \gamma(\mu + \beta)
\end{equation}
\noindent
and
\begin{equation}
{d\mu \over ds} = \gamma(1 -\mu^2) \left[ { 1+\beta\mu \over r} - \Lambda \right] \,\,\,.
\end{equation}

We choose N$_c$ rays to strike the stellar core, while the remaining rays are set by the impact parameter ``$p$" which is defined by the radius grid. To obtain the curved characteristics we integrate the above equations using a Runge-Kutta technique. The accuracy of the integration can be checked by using the relation between $\mu_s$ and $\mu_o$, and noting that $\mu_s$ is simply
$\sqrt{[1-(p/r)^2]}$.

As an aside we note that the $\partial \beta /\partial t$ term may be written as $D\beta/Dt - v \partial \beta /\partial r$. Thus for a Hubble flow  $\Lambda$ reduces to
\begin{equation}
\Lambda=\mu {\partial \beta \over \partial r} \;\; .
\end{equation}

Along a characteristic ray the transfer equation reduces to
\begin{equation}
{\partial I \over \partial s} - \nu \Pi  {\partial I \over \partial \nu} = \eta -\left(\chi + 3 \Pi \right) I
\end{equation}

\noindent
with
 \begin{equation}
 \Pi= \gamma \left[ {\beta (1-\mu^2) \over r} + \mu \Lambda \right]
 \end{equation}

\noindent
Using backward differencing in frequency, and using $k$ to denote the current frequency,  we have
\begin{equation}
{dI_k \over d s}  =  \eta_k+ a_{k}I_{k-1}   -\left (\chi_k + 3\Pi +a_k \right) I_k
\label{Eqn_ray}
\end{equation}

\noindent
with
\begin{equation}
a_k = { \nu_k \over \nu_{k-1}-\nu_k} \Pi
\end{equation}

\noindent
This may alternatively be written as
\begin{equation}
{dI_k \over d \tilde \tau_k}  =  I_k - \tilde S_k
\label{Eqn_ray}
\end{equation}

\noindent
with
\begin{equation}
d \tilde \tau_k=-\left (\chi_k + 3\Pi +a_k \right) ds
\end{equation}
\noindent
 and
\begin{equation}
\tilde S_k= (\eta_k + a_{k}I_{k-1}) / \left (\chi_k + 3\Pi +a_k \right)
\end{equation}
Since \eq~\ref{Eqn_ray} is the standard form of the transfer equation along a ray it can be solved by the usual techniques. Our approach is to assume that $S_k$ can be represented by a parabola in $\tau_k$ over three successive grid points, and we analytically integrate the transfer equation using this representation. We first integrate inwards to obtain $I_\nu^-$, then outwards to obtain $I_\nu^+$. The radiation moments are obtained by numerical quadrature assuming $I_\nu$ is a linear function of $\mu$ --- this is preferred over higher accuracy schemes because of stability.

More recently an alternative approach has been suggested. In this mixed frame approach the transfer equation is written in terms of an affine parameter, and the comoving frequency (or wavelength; 
\citealt{CKB07_rad,BHC09_3Drad}). Its advantage is that the transfer equation is solved along  geodesics (straight lines in our models) allowing the approach to be easily generalized to three dimensions. Like the method of \cite{Mih80_rel_flow}, the opacities and emissivities are evaluated in the comoving-frame.

\section{Time Dependent Rate and Radiative Equilibrium Equations}
\label{Sec_stat}

The time-dependent rate equations can be written in the form

    \begin{equation}
      \rho {D n_i/\rho \over Dt} = {1 \over r^3} {D(r^3 n_i)  \over Dt} =  \sum_j \left(n_j R_{ji}  - n_i R_{ij}\right)\,,
      \label{eq_SSE}
    \end{equation}

\noindent
where $\rho$ is the mass density, $n_i$ is the population density of state $i$ for some species, and $R_{ji}$ is the rate from state $j$ to state $i$.  In general, the $R_{ji}$ are functions of $n_e$ (the electron density), temperature, and the radiation field \citep[see, e.g.,][]{Mih78_book},and when charge exchange reactions are included they are also dependent on $n_i$. Thus the equations are, in general, explicitly non-linear, although with the exception of the temperature, this non-linearity is only quadratic. Of course, there is a strong implicit non-linearity resulting from the dependence of the radiation field on the populations. In {\sc cmfgen} we allow for the following processes: bound-bound processes (including two photon decay in H and He\,{\sc i}), bound-free processes, collisional ionization and recombination, collisional excitation and de-excitation, charge exchange (primarily with H\one\ and He\,{\sc i}),  and 
Auger ionization \citep{Hil87_A,HM98_blank}.

 The energy equation has the form
    \begin{equation}
    \rho {De \over Dt} - {P \over \rho} {D\rho \over Dt}= 4\pi \int d\nu (\chi_\nu J_\nu  - \eta_\nu) + \dot \epsilon_{\rm decay} \,,
     \label{eq_energy}
     \end{equation}

\noindent where $e$ is the internal energy per unit mass, $P$ is the gas pressure, and
$\dot \epsilon_{\rm decay}$ is the energy absorbed per second per unit volume. When gamma-ray deposition is
assumed to be local (see Appendix), it is simply the energy emitted  by the decay of unstable nuclei (excluding the energy carried off by neutrinos). To improve numerical conditioning, continuum scattering terms are explicitly omitted from the absorption and emission
coefficient in equation~\ref{eq_energy}. $e$ can be written in the form

      \begin{equation}
      e = e_K +e_I\,,
      \end{equation}

\noindent
where

      \begin{equation}
       e_K={3 kT(n+n_e) \over 2 \mu m n }\,,
      \end{equation}

\noindent
and
      \begin{equation}
      e_I=  \sum_i {n_i E_i \over \mu m n}\,.
      \end{equation}

\noindent
In the above, $n$ is the total particle density (excluding electrons), $m$ is the atomic
mass unit, $\mu$ the mean atomic weight,  $T$ is the gas temperature, and $E_i$ is the total 
energy (excitation and ionization) of  state $i$.

The influence of the time dependent terms in the rate and radiative equilibrium equations on ejecta properties and synthetic spectra has previously been discussed by \cite{DH08_time}.  They found that these terms are crucial for explaining the strength of the H\,{\sc i} lines in Type II-P spectra during the recombination epoch. Importantly, the terms also increase the He\,{\sc i} line strengths in early spectra so that their line strengths are in better agreement with observation. However the terms do not just influence H and He lines --- other lines are also affected since the electron density changes \citep{DH08_time}. As noted earlier, \cite{UC05_time_dep}, showed that
the time dependent terms were also important for modeling the H\one\ lines in the Type II-peculiar SN 1987A. More detailed work on the influence of time-dependence on Type IIP SN spectra has been presented by \cite{DH10_time}.

\noindent
\section{Deposition of Radioactive Energy}
\label{Sec_rad_energy}


During a SN explosion radioactive elements are created (for a review, see, e.g.,  \citealt{Arn96_book}). Type II SNe produce between 0.01 
and 0.3\,\Msun\ of $^{56}$Ni \cite[e.g.,][]{Ham03_SNII_rev} while a Type Ia SN can eject anywhere between 0.1 to 1.0\,\Msun\ of \isoni\ \cite[e.g.,][]{HK96_Ia_lc,SLW06_Ni_masses}.  The influence of the radioactive material on the light curve (and spectrum) very much depends on the $^{56}$Ni mass and distribution, the size of the progenitor, and the ejecta mass. Radioactive decay powers the light curve of all SN types during the nebular phase (excluding those powered by an external source such as a magnetar).  The decay of $^{56}$Ni (and its daughter nucleus, $^{56}$Co) powers the entire light curve of Type Ia SNe. For SN 1987A, associated with the explosion of a BSG progenitor star,  $^{56}$Ni heating begins to become 
apparent at approximately day 10, and is responsible for the broad maximum, which lasts for approximately 100 days \citep{SN90_SN1987A_lc,BLB00_SN87A_lc}.




The most important nuclear decays are \\
\begin{eqnarray}
^{56}\hbox{Ni} \rightarrow\   ^{56}\hbox{Co} + \gamma  &&  (\hbox{half life} = 6.075\, \hbox{days}, \nonumber \\
                  &&  \Delta E = 1.7183\,\hbox{MeV per decay})   \nonumber
\end{eqnarray}
\noindent
and
\begin{eqnarray}
^{56}\hbox{Co} \rightarrow\   ^{56}\hbox{Fe} + && \hspace{-0.25in} \gamma + \hbox{e}^+  
                ( \hbox{half life} = 77.23\, \hbox{days}, \nonumber \\
                 && \Delta E (\hbox{\grays}) =  3.633\,\hbox{MeV per decay}; \nonumber \\
                 && \Delta E (\hbox{positrons}) =  0.1159\,\hbox{MeV per decay}).  \nonumber
\end{eqnarray}

\noindent

The energies and lifetimes are all from http://www.nndc.bnl.gov/chart/, which for \isoni\ and \isoco\ are
based on the work of \citet{HHZ87_nuc}. 
Energy release in the form of neutrinos is neglected, since ejecta densities at the times of interest are all
well below the neutrino-trapping densities.


\noindent
To implement the decay of the radioactive elements we analytically solved the 2-step decay chains. As a consequence, the accuracy of time-dependent abundances is not affected by our choice of time-step. Another corollary of our approach is that we do not use the instantaneous rate of energy deposition  -- rather we use the average rate of deposition over the time step. In practice, the difference between
the two values is small.  Our nuclear-decay reactions are currently being updated so that we can accurately follow the evolution of the SN ejecta abundances.

In Type II SNe it is reasonable to assume that all the energy, emitted as \grays\ or positrons, is deposited locally. At later times (but well after maximum brightness) this assumption breaks down, and \grays\ can travel long distances depositing energy elsewhere in the SN, or escaping entirely. On the other hand, in Type I SN, the small ejecta masses mean that non-local \gray\ energy  deposition, and \gray\ escape, especially in models with extensive mixing, needs to be taken into account even at relatively 
early times {\cite[e.g.,][]{HKM92_gam_rays}. We have developed a \gray\ transport code (see Appendix),  to allow for non-local \gray\ energy deposition.

At present we simply assume that the energy from radioactive decays goes into the thermal pool. This is likely to be valid for modeling the early light curve of most Type II SNe (but well beyond maximum brightness) but may fail relatively early for other SNe\footnote{Type II SNe have much larger ejecta masses than do Type I SNe. As a consequence radioactive material is generally hidden from view for much longer. Deep below the photosphere, conditions are highly ionized and close to LTE, and any radioactive energy deposited will be thermalized.}. One factor limiting the influence of non-thermal processes at earlier times is the mass of the ejecta \citep{DHLW12_SNIbc}. Type IIP SN have much greater ejecta masses than do most Type I SN, and hence the photosphere at early times is confined to the outer layers where $^{56}$Ni has not been mixed.} Non-thermal ionization and excitation processes, for example, are believed to be important  for explaining the appearance of He\,{\sc i} in SNe Ib \citep{L91_HeI,DHL11_Ibc}, and for the production of H\one\ lines in the nebular phase of Type II SNe, and in SN 1987A  \citep{XM91_87A_energetic,KF92_gam_rays,KF98_1987A,KF98B_1987A}.  Non-thermal excitation and non-thermal processes have recently been incorporated into {\sc cmfgen} \citep{LHD12_NT} and we are currently investigating their influence on the spectra of SN 1987A \citep{LHD12_NT}, Type Ibc SNe \citep{DHLW12_SNIbc}, and Type Ia SNe \citep{DH12_SNIa}.

\section{Numerical Solution}
\label{Sec_num_sol}

To handle the Lagrangian derivative we adopt an implicit approach \citep{HMK93_nm, SMN92_Zeua} which is explicitly stable, for both the \rte, the rate equations, and the radiative energy balance equation. Thus we write
  \begin{equation}
     \left({DX \over dt}\right)_i = { X_i - X_{i-1} \over t_i - t_{i-1}}
   \end{equation}

\noindent
where $i$ refers to the current time step, and $i-1$ refers to the previous time step.  As our approach is only first order accurate in time, the time step must be kept ``small". An explicit approach cannot be used as the conditions on the time step would be unnecessarily restrictive \citep[e.g.,][]{SMN92_Zeua}.

All other terms in the \rte\ and rate equations are evaluated at the current time step. Since the quantities at the previous time step are known, the Lagrangian derivative simply introduces extra source terms into the equations. With the above scheme the zero-order moment equation, for example, becomes
  \begin{eqnarray}
     \left( {1 \over r^2} {\partial r^2 H_\nu  \over \partial r} \right)_i
  &-&  \left( {\nu v \over rc} { \partial J_\nu \over \partial \nu } \right)_i  \nonumber \\
   &=& (\eta_\nu)_{i} + \left( {r_{i-1}\over r_i} \right)^3 {J_{\nu , i-1} \over c \Delta t} \nonumber \\
   &-& \left( \chi_\nu + {1 \over c \Delta t}\right)_i  J_{\nu ,i} \,\, .
 \label{eq_zero_mom_diff}
 \end{eqnarray}

\noindent
This has exactly the same form as the equation normally solved  in {\sc cmfgen}, and can be differenced in the usual way \citep{MKH75_I, MKH76_III}. The only difference is that there is a modification to the opacity and emissivity arising from the radiation field at the previous frequency. The solution of the rate equations has previously been discussed by \cite{DH08_time}.

 An alternative approach would be to choose a semi-implicit approach with all terms evaluated at the midpoints of the time step. This would be second order accurate, and should also be stable. Its disadvantage is an increase in complexity --- additional terms would need to be saved from the previous time step.  \cite{HMK93_nm} use an adjustable differencing approach, controlled by a parameter $\theta$, which allows for an implicit approach, a semi-implicit approach, or a fully explicit approach. Another possibility, utilized in some evolutionary codes and by \cite{L05_MC}, is to generate a difference formula using the two previous time steps.
 
At depth the opacity and emissivity are large, and generally the time terms in the ``modified opacity'' and ``modified emissivity'' are almost negligible (unless a very small time step is used), and their influence on the radiation transfer at a given frequency could be neglected if one were only interested in the formal solution of the transfer equation. However the terms are of crucial importance for the energy balance, since their importance is enhanced when we integrate the transfer equation over frequency (\S\ref{Sec_gray}).  In particular, if we assume radiative equilibrium holds the opacity and emissivity are ``analytically'' cancelled from the integrated transfer equation, and thus the time dependent terms can be very important. As can be seen from the gray transfer equations (\S\ref{Sec_gray}), the temperature at large optical depths is almost set entirely by the time-dependent term, and for a homologous flow without 
energy deposition from radioactive decay it simply scales as $1/t$.

For most of our calculations we have set the time step to be 10\% of the current SN age (Section~\ref{sect_time}). At some epochs, a smaller time step may be needed. Since we need to iteratively solve the \rtes\ in conjunction with the rate equations, a factor of 2 reduction in the time step does not mean a factor of 2 extra computational effort, since each time step will require fewer iterations (because the starting solution will be closer to the final solution).

As we are considering SN ejecta in homologous expansion, we use the velocity as our Lagrangian coordinate.

Before beginning our simulations, the hydrodynamical models are mapped onto a revised grid optimized for the computation of the radiation transfer. The new grid is defined in terms of Rosseland optical depth, and generally we require at least 5 points per decade of optical depth. Similar grids are generally used in the modeling of stellar atmospheres, and are preferred to a grid defined in mass and optimized for the hydrodynamic calculations. At present we generally perform a new mapping at each time step (although the mapping is done using the calculations at the previous time step rather than a hydrodynamical model). Two problems can potentially arise with our mapping scheme:
\begin{enumerate}
\item
In standard 1D SN models there are often rapid changes in spatial composition and density distribution. Due to the numerous re-mappings, these rapid changes 
suffer from numerical diffusion in space. At present we do not make any special attempt to limit this diffusion. Numerical diffusion is less of a problem in mixed models 
because the spatial gradients are smaller. The steep density gradients are also stronger in core-collapse SN ejecta because of the chemically-stratified progenitor 
envelope structure.
\item
In some models (particularly for  SN 1987A) ionization fronts can arise. Because the optical depth (e.g., in the Lyman continuum) can vary rapidly  across the front, complications arise in the radiative transfer and in model convergence. To overcome these problems we semi-automatically revise the grid across the front. Crudely, we adjust the grid so the front is better resolved. A control file, which can be altered while {\sc cmfgen} is running, is used to specify parameters for adjusting the grid. If necessary, the grid can be adjusted at each iteration.
\end{enumerate}

\section{Linearization}
\label{Sec_linearization}

The transfer, energy, and rate equations are a coupled set of non-linear equations. Thus their simultaneous solution requires an iterative technique. As described by \cite{H90_lin} we adopt a linearization approach. While cumbersome to implement, we have found this technique works extremely well for the modeling of hot stars and their stellar winds. Another common approach is to utilize approximate lambda operators to allow for the coupling of the level populations with the radiation field \cite[e.g.,][]{Wer86_ALO,LH95_TLUSTY,HB99_phoenix}. In some cases this is combined with other approaches ---  \cite{Hof03_ALI} also utilizes an equivalent-two level-atom  approach in his SN modeling.

The linearization of the \rte\ and the rate equations is straightforward, and as it is very similar to that presented by \cite{H90_lin} it will be described only briefly. In our approach we linearize both the rate equations and 
the \rte. Using the linearized \rte\ we can solve for the $\delta J_\nu$ in terms of $\delta \chi_\nu$ and $\delta \eta_\nu$, and hence in terms of $\delta n_1, \delta n_2, \delta n_3, \dots \delta N_e$, 
and $\delta T$.  In principle, the $\delta J_\nu$ depend on the perturbed populations at every depth, however we only allow for the coupling  locally (diagonal operator) or between adjacent depths (tridiagonal operator). Using these equations we eliminate the radiation field ($\delta J_\nu$) from the linearized rate equations. For the diagonal operator we obtain a set of \NV\ simultaneous equations (where \NV\ is the total number of variables) at each depth. For the tridiagonal operator we obtain a block tridiagonal system of equations ($\ND \times \NV$ in total, where \ND\ is the number of grid points). The latter can be efficiently solved using a block-matrix implementation of the Thomas algorithm for Gaussian elimination. In a typical SN model $\ND=100$ and $\NV\sim 2000$, yielding 200\,000 equations. The 2000 super-levels typically represent 5000 to 10\,000 atomic levels.

Starting estimates for the temperature and level populations are obtained as follows. For the temperature structure we either adopt a scaled version of the gray temperature structure (see \S\ref{Sec_gray}), or simply adopt the temperature from the previous solution. For the level populations we adopt the departure coefficient from the previous time step, although we also have the option to adopt LTE values (useful only for the first model of a time sequence).


 To solve the linearize rate equations we use LU decomposition followed by backward substitution utilizing standard LAPACK\footnote{LAPACK is a software library for numerical linear algebra. Documentation and the routines are available at http://www.netlib.org/lapack/} routines.
 As discussed by \cite{H03_sol_see}, we improve stability by doing the following:
 \begin{enumerate}
 \item
  We scale the simultaneous equations so that we solve for the fractional correction $\delta n_i/n_i$.
 \item
 We use the LAPACK routine  {\sc dgeequ}  to pre-condition the matrix.
 \end{enumerate}
These two procedures work very well, and we (generally) have no difficulty in solving systems of equations with \NV$=2000$. When problems do occur it is usually because of poor starting estimates.

The advantage of the diagonal operator is that the construction of the linearization matrix is faster; each depth can be treated independently; in the early stages of the iteration procedure convergence can be more stable; and it requires less memory (nominally down by a factor of 3 but because of our implementation we save only a factor of 2). The advantage of the tridiagonal operator is its speed of convergence --- far fewer iterations are needed to reach a specific convergence criterion. With both operators, Ng acceleration \citep{Ng74,Auer87_Ng} can be used to accelerate the convergence.

The time taken per iteration is very model dependent, and depends on the type of iteration.
Iteration steps (for \ND=125, \NV=2319, $\sim$400,000 bound-bound transitions) in which the linearization matrix is held fixed take about  $\sim$15 minutes, $\Lambda-$iterations take somewhat longer ($\sim$20 minutes) while a full iteration may take 4 hours. The computations
were formed using 4 processors (Quad-Core AMD Opteron[tm] Processor 8378; 800 MHz), and
the computational times are about a factor of 2 longer than that obtained using a Mac Pro. The computational time scales as $\ND^2$.

\section{Super-levels}
\label{Sec_SLs}

To reduce the complexity of the non-LTE problem it is usual to use super-levels (SLs), or a variant thereof,  to reduce the number of variables, and hence the number of rate equations that need to be solved \cite[e.g.,][]{And89_Sls,DW93_SLs, HL95_TLUSTY, HM98_blank}.  In the procedure adopted by \cite{HM98_blank}, SLs are used as a means of reducing the number of level populations that need to be determined, while all transitions are still treated at their correct wavelength. Within a SL we assume that the states have the same departure from LTE (although more complicated assumptions can also be adopted); thus this procedure can be considered to be the natural extension of LTE.

However SLs can introduce inconsistencies which may adversely affect convergence. Consider a three level atom, with two excited levels which are relatively close in energy, and which are treated as a single SL in the rate equations.

The rate equation for the SL contains terms of the form
  \begin{equation}
    n_2 A_{21} Z_{21} + n_3 A_{31} Z_{31} \nonumber
  \end{equation}
\noindent
while the radiative equilibrium equation will contain terms of the form
\begin{equation}
h\, \nu_{21} n_2  A_{21} Z_{21} + h \,\nu_{31} n_3 A_{31} Z_{31} \nonumber
\end{equation}
where $Z_{ij}$ is the net radiative bracket, $A_{ij}$ is the Einstein A coefficient, $h$ is Planck's constant, and
$\nu_{ij}$ is the transition frequency. When levels 2 and 3 are part of the same SL, and if the bound-bound transitions are the dominant transitions determining the population of the SL and are a significant coolant, an inconsistency arises since the ratio $n_2/n_3$ is fixed by the assumption that two levels have the same departure from LTE. In extreme cases this inconsistency can cause models not to converge.

As noted by \cite{H03_sol_see}, another manifestation of this inconsistency is that when radiative equilibrium is obtained the electron-energy balance equation will not be satisfied. The level of inconsistency gives an indication of the errors introduced by the use of SLs. To overcome the inconsistency, \cite{H03_sol_see} simply replaced $\nu_{31}$ and $\nu_{21}$ by an average frequency, $\bar \nu$ in the radiative equilibrium equation. This procedure overcame convergence difficulties and yielded better consistency of the electron-energy balance equation with the radiative equilibrium equation. Detailed tests, performed by splitting the important SLs, confirmed the validity of the procedure in the atmosphere/winds of massive stars. At high densities the scaling approximation can be switched off.

In SNe the simple scaling described above has potential difficulties. Firstly, the lower densities and higher velocities increase the importance of line scattering compared with continuous processes. Second, the radiative equilibrium equation appears on the  right-hand side of the zeroth moment of the gray equation. Since the above scaling is not done in the transfer equation there is an inconsistency between the transfer equation and the energy balance equation. In the present case the assumption of radiative equilibrium is untrue, due to the work done on the gas, and the deposition of energy from nuclear decays, but there will still be an inconsistency between the transfer equation and energy balance equation.

To remove the inconsistency, but at the same time retain the benefits of the SL approach, we now simply scale the line opacity and emissivity.  The emissivity and opacity due to the transition {1--3} is scaled by $\bar \nu / \nu_{13}$  -- it is necessary to scale both the emissivity and opacity so that $B=\eta/\chi$ at depth. Most of the scalings are less than 10\%, and thus of the same order of  accuracy as metal line transition probabilities. In principle the influence of the assumption is easily tested by altering the SL assignments --
in practice this can be difficult because of memory requirements. Alternatively, after the temperature has converged, the scaling can be switched off, the temperature can be held fixed, and the populations can be updated. With this technique the direct influence of the scaling on the observed spectrum can be removed. No scaling is used when computing observed spectra with {\sc cmf\_flux} (\S\ref{Sec_comp_obs_spec})\footnote{As noted by the referee it is possible to define an average A and Z value for a transition between two super-levels. In practice this would
remove the inconsistency when the radiative equilibrium equation is defined in terms of the net rates although in practice it will yield results similar to our frequency scaling procedure. However this procedure does not solve the problem with the radiative equilibrium equation on the RHS of
the transfer equation where there is no distinction between lines and continuum.}.

\section{Convergence}
\label{Sec_convergence}

In type II SNe we expect the temperature at large optical depths to evolve adiabatically\footnote{Radioactive decay will modify this simple scaling.} as $1/r$ from the initial configuration, and as this is a local scaling we expected convergence to be achieved despite the extremely high optical depths. However, initial models failed to converge, with the temperature stabilizing at the gray temperature adopted for the initial estimate. Convergence was only an issue with the temperature structure. When the temperature was held fixed, populations would converge rapidly at all depths.

The lack of convergence was traced to two issues, and both are inherently linked to the large optical depths. Because of the large optical depths, any term included in the linearization of the energy balance equation must be accurately linearized in order that the correct cancellation can occur.

Consider the following contribution, by a single line at a single frequency, to the radiative energy balance equation.
\begin{eqnarray}
\Delta RE &=& 4\pi \left(\chi_c + \phi \chi_l \right)\left(J - S\right) \\
                &=&4\pi \left(\chi_c + \phi \chi_l \right)J - \left( \eta_c + \phi \eta_l \right)
\end{eqnarray}

\noindent
In the above, the subscripts ``$c$" \& ``$l$" are used to denote the continuous and line contributions, and $\phi$ is the
intrinsic line emission/absorption profile. As the medium is optically thick we have $J=S$ and hence
\begin{eqnarray}
J&=& {\eta_c + \phi \eta_l \over \chi_c + \phi \chi_l} \\
  &=&  {S_c + \theta \phi S_l \over 1 + \theta \phi}
\end{eqnarray}

\noindent
with $\theta= \chi_l/\chi_c$. Obviously $\chi(J-S)=0$. However, if $S_c \ne S_l$ the line and continuum term will
make contributions that are identical in magnitude, but of opposite sign. This cancellation must also
be maintained in the linearization. The continuum contribution $\chi_c(J-S_c)$ to $\chi(J-S)$ is, for example,
\begin{equation}
{\phi(\eta_l-\theta \eta_c) \over 1+\theta \phi} \,.\nonumber
\end{equation}

In our original version of {\sc cmfgen} we integrated the line contribution to the linearization over the full line before adding it to the linearized energy balance equation. For simplicity, and computational requirements, we added the linearized continuum contribution to this term using data at the final frequency in the line. However, as the continuum varies across the line, full cancellation was not achieved. Since cancellation was not achieved we effectively overestimated the effect of any change in the temperature on the energy balance. This, in turn, meant that the corrections needed to achieve balance were severely underestimated, and hence stabilization of the temperature structure occurred. 
The problem was solved by performing the continuum linearization at each frequency in the line (although this is only necessary, and is only done, for the energy balance equation).

A similar problem arose from the use of impurity levels. The concept of impurity levels was introduced into {\sc {\sc cmfgen}} to minimize the number of levels that were fully linearized. The levels are linearized for rates associated with their own species, but their influence on other species is neglected in the linearization. The use of impurity levels does not affect the solution; it only has an influence on the convergence of the code. In our implementation, exact cancellation was not achieved for terms involving impurity levels, and the rate of convergence was severely curtailed in the inner region.

\begin{figure}
\epsfig{file=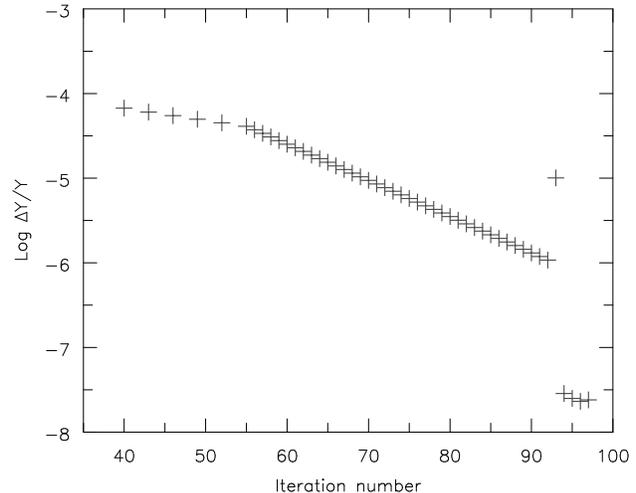,width=6.5cm,angle=-90}
\caption[]{Illustration of the corrections to the temperature at a depth $\tau_{\hbox{Ross}}=1.5 \times 10^6$ in a Type II SN model. The model, at time 2.3\,d after the explosion, is similar to those used to model SN 1987A by  \cite{DH08_time, DH10_time}. The convergence shown is fairly typical of the convergence behavior at depth. The temperature correction for iterations 1 to 39 is zero. All models begin with $\Lambda$ iterations until the corrections stabilize, and then a mixture of lambda and non-$\Lambda$ iterations with T fixed at all depths is performed. An Ng acceleration was performed after iteration 93 (hence the spike at iteration 94).
}
\label{fig_conv1}
\end{figure}

\begin{figure*}
\epsfig{file=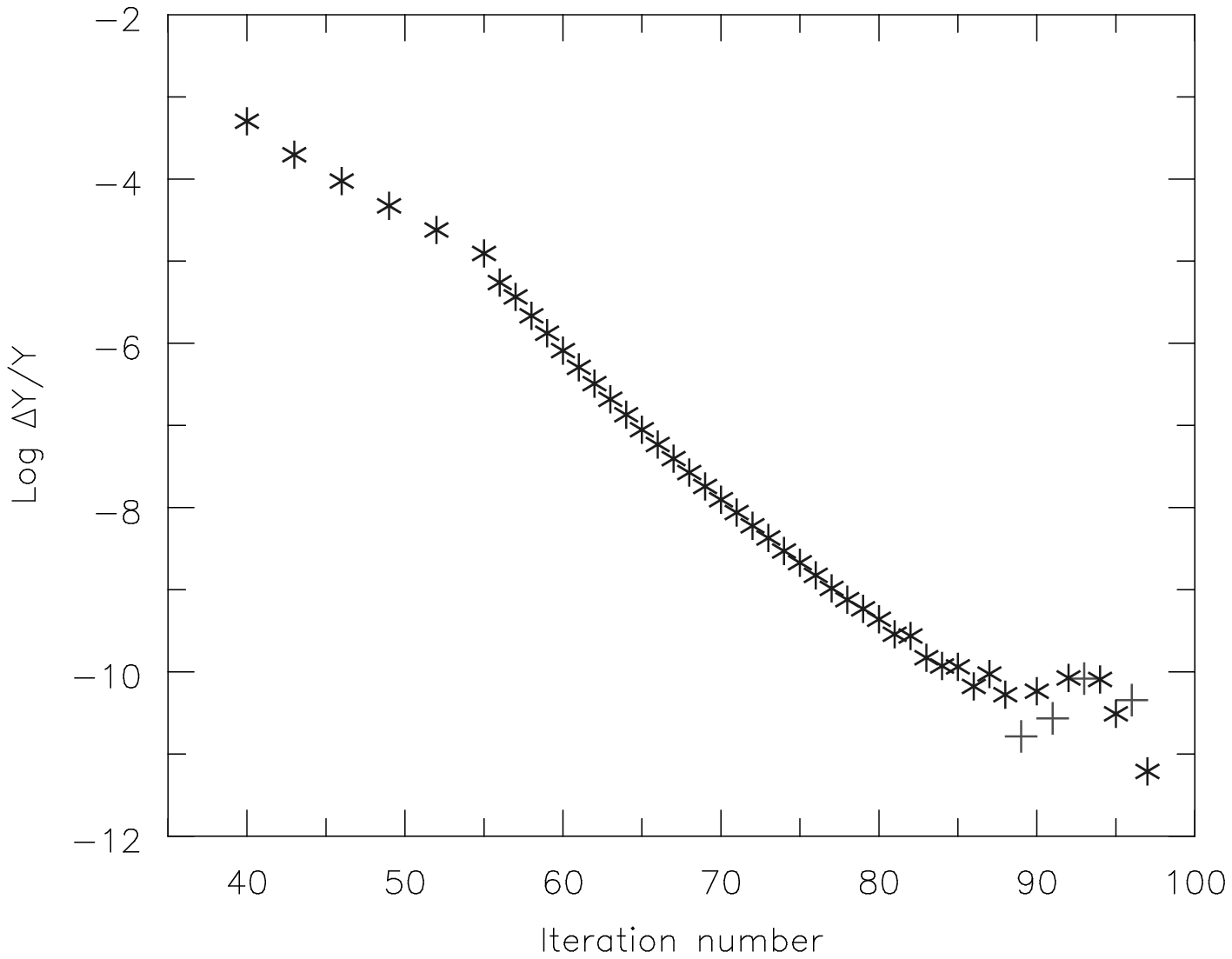,width=4.5cm, angle=-90}
\epsfig{file=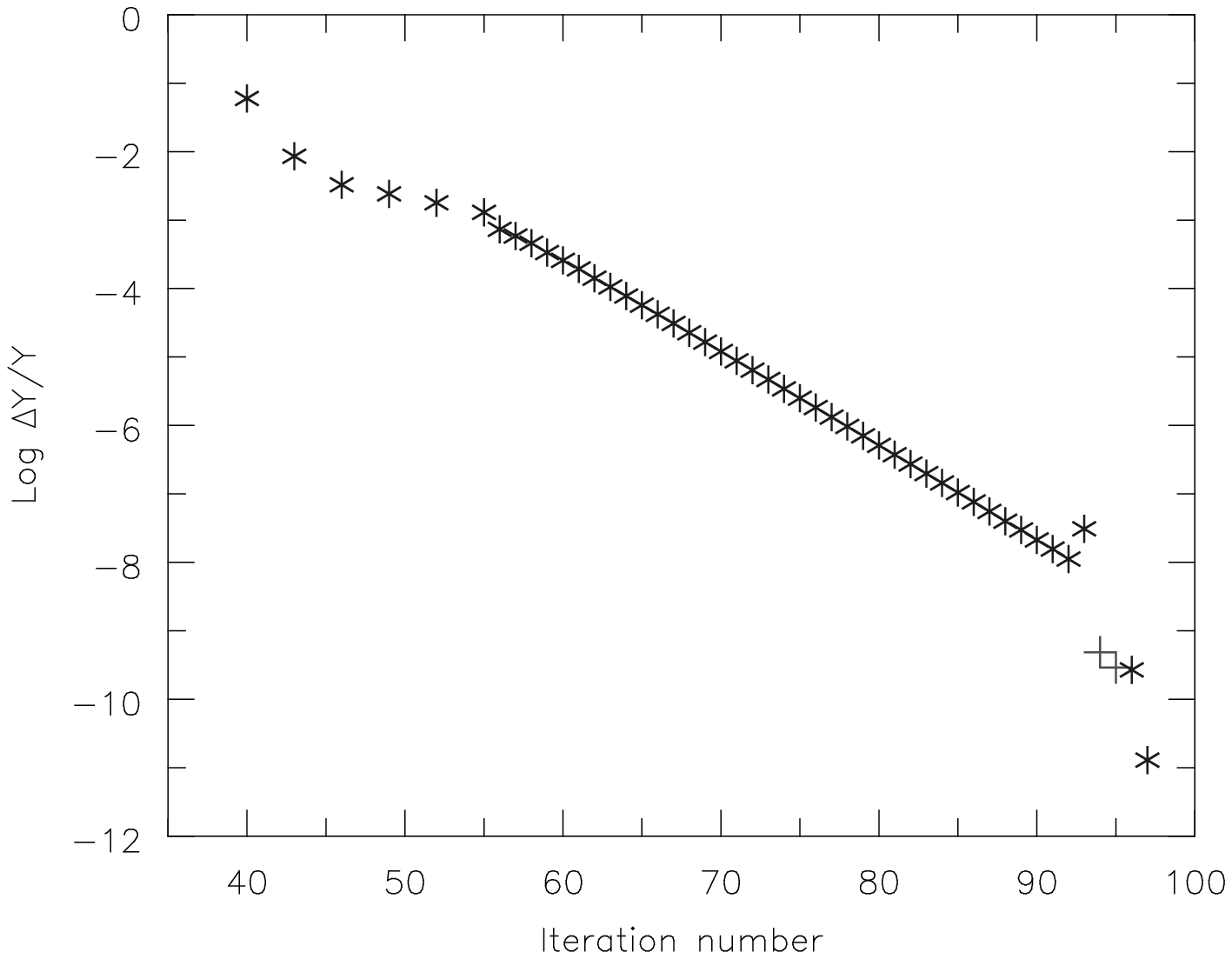,width=4.5cm, angle=-90}
\epsfig{file=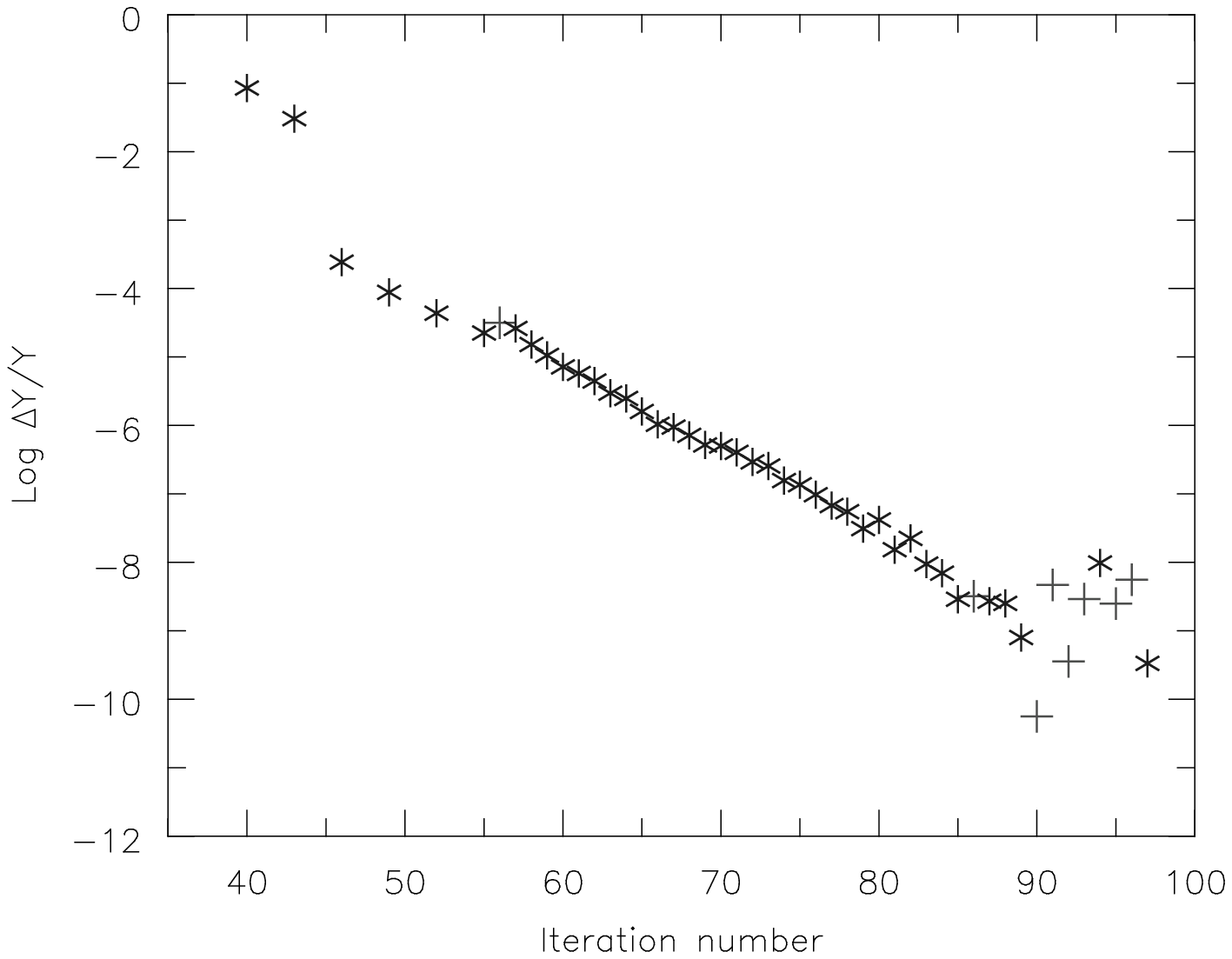,width=4.5cm, angle=-90}
\caption[]{Same as \fig~\ref{fig_conv1}, but now showing the convergence at a depth $\tau_{\hbox{Ross}}$ of  $\sim$\,1000 (left), $\sim$\,1 (middle), and $\sim$\,0.001 (right). The two types of symbols are used to indicate the sign of the correction.}
\label{fig_conv2}
\end{figure*}

To illustrate the convergent properties of the models we illustrate  in \figs~\ref{fig_conv1}--\ref{fig_conv2} the convergence properties of a typical Type II SN model. Only the temperature is shown. In general, it is the temperature that determines the global convergence properties of the model --- when the temperature is held fixed convergence is usually rapid. For the model shown convergence is achieved in 100 full iterations, with an iteration defined as any step which changes the populations (e.g., a $\Lambda$ iteration, a full iteration, or an Ng acceleration).

While the temperature corrections in Fig.~\ref{fig_conv1} are small, they help
to illustrate that our linearization procedure does achieve convergence at large optical depth.
As noted above, in early work temperature corrections stabilised, and it was unclear whether convergence was being achieved. When we use the gray temperature solution at depth for
the initial temperature estimate, the predicted temperature corrections are small. However, when
we use the temperature from the previous time step, the error in T is of order 10\% (for a 10\% time step)
and we find good convergence provided we keep the populations consistent with the
current temperature structure. It also should be noted that the actual error achieved at a given iteration depends on both the size of the correction, and the rate at which the correction is changing. In Fig.~1, the true error, as estimated over iterations 55 to 60, is approximately a factor of 10 larger than the current correction. In Type II SN it is the sharp H ionization front, which moves to smaller velocities as the SN ages, that controls convergence of the model.

Corrections to the populations are generally much larger than the temperature corrections, and tend to behave more erratically, especially in the early iterations. Experience has allowed us to arrive at the following iterative procedure:

\begin{enumerate}

\item
 Perform $\Lambda$ iterations until the maximum correction is less than 50\%.
The temperature is held fixed.

\item
Perform full iterations, with the temperature held fixed, until the
maximum correction is less than 50\%. If the maximum correction is greater than 200\%, we revert to $\Lambda$ iterations.

\item
When the maximum correction on a full iteration is again less than 50\%, allow the temperature to vary. If the maximum correction is greater than 200\%, we revert to $\Lambda$ iterations. When we are no longer switching between full and $\Lambda$ iterations, we only compute the variation matrix (i.e., the linearized rate equations) every third iteration.

\item
When the maximum correction is less than $\sim$\,$x$\% ($x$ is an adjustable parameter typically set to 5) we hold the variation matrix fixed. It is recomputed if the maximum correction becomes larger than $3\times x$\%.

\item
When 4 successive iterations are less than VAL\_DO\_NG (typically a few percent) we perform an Ng acceleration. We repeat Ng accelerations every 10 to 20 iterations.

\end{enumerate}

This procedure is automatically handled by {\sc cmfgen}, although it is also possible to manually force a different behavior by restarting the code. Many of the control parameters are adjustable, although in practice we tend not to change them very much.

In hot-star models we also have the option to fix the temperature automatically in the outer region until the populations have stabilized. This generally works well in hot-star winds, but in SN models it caused convergence problems around ionization fronts.

A typical model is stopped when the maximum convergence is less than 0.1\%, although most levels have already achieved a higher convergence. Several output files can be checked to verify that true convergence has been achieved. Some models in a sequence are converged to a much higher accuracy (e.g., 0.001\%) to further prove convergence. Convergence to higher accuracy is generally not time consuming since the variation matrix can generally be held fixed --- it is the initial iterations, with a mixture of $\Lambda$ and full iterations, that take the most time.

In Type II SN models two factors can affect the speed of convergence.

(1) At a few depths (often only a single depth), a few levels may oscillate and fail to converge. These levels are generally unimportant, and do not affect the spectrum. However, they can affect convergence since it is the maximum change that controls the iteration procedure and ultimately determines the convergence. NG accelerations, averaging the last two full iterations, and forcing multiple $\Lambda$-iterations, can overcome the problems. More recently we have implemented a procedure to use negative relaxation at the problem depths -- that is, we scale all corrections at those depths by a scale factor (typically 0.3). This procedure is often successful at accelerating convergence.

(2) Ionizations fronts: These tend to show rapid changes over a few successive points. The position of the front  oscillates, and there is often a strong coupling between the front location and the temperature. In these cases the most reliable technique to facilitate convergence is
grid refinement in the neighborhood of the front. This is done semi-automatically in {\sc cmfgen} --- a parameter file (which can be changed during the run)  controls if, when and how the refinement is done.

\section{Testing}
\label{Sec_testing}

With a code as complicated as {\sc cmfgen}  there is no easy way to fully test the code. However numerous checks
have been performed to test individual components. For example, solutions obtained with different transfer
modules have been compared, and found to agree (in limiting cases such as small $v/c$). At depth (e.g., $\tau > 10$), the solution 
to the gray and non-gray transfer equations also agree. Light curves of Type II SN computed with {\sc cmfgen} show reasonable 
agreement to those computed using the hydrodynamics code {\sc v1d} \citep{livne_93,DH10_time}. A strong check, although not definitive, is 
also provided by our modeling of  SN 1987A. The time-sequence of models  for SN 1987A, begun at 0.3\,d, shows a remarkable 
agreement to observations of  SN 1987A \citep{DH10_time}  --- remarkable since there were no adjustable parameters in the modeling.

Additional checks are provided by auxiliary files which list and check various processes and rates such
as the ionization balance of all species and the electron heating/cooling balance modified for advection, adiabatic cooling and nuclear energy deposition. As noted by \cite{H03_sol_see}, the later equation, in the absence of SLs, is a linear combination of the radiative equilibrium equation and rate equations.
In the presence of SLs this equality no longer holds, and thus the electron energy balance equation provides a check on possible problems with SL assignments.

\section{Computation of Observed Spectra}
\label{Sec_comp_obs_spec}

The computation of the observed spectrum can be done in three ways:

\begin{enumerate}
\item
Compute the observed spectrum using a Lorentz transformation of the comoving-frame boundary intensities computed by {\sc cmfgen} (Section~\ref{CMF_OS}).
\item
Compute the observed spectrum using the comoving-frame boundary intensities computed using a separate calculation with {\sc cmf\_flux}. The advantage of this technique over the {\sc cmfgen} calculation is that we can use a finer spatial and frequency grid to improve the accuracy of the computation, we include all spectral lines for the adopted model atoms, and we use the technique of \cite{RH94_es} to compute the electron-scattering source function. In this technique the redistribution of photons in frequency space due to the thermal motions of the electrons is explicitly treated.  The redistribution by the bulk motion of the gas is automatically allowed for by performing the radiative transfer calculations in the comoving frame. In the computation of the level populations (i.e., in {\sc cmfgen}) we generally treat the scattering coherently in the comoving-frame -- that is, we only allow for redistribution caused by the bulk motion of the gas. With the coherent scattering approximation the radiation is not explicitly coupled to that at other frequencies by a complex weighting function, and hence convergence is more stable.

\item
Using the emissivities and opacities computed in the comoving frame, compute the observed spectrum using an observer's frame calculation. This calculation has a higher accuracy than the calculation done in the comoving frame. This occurs since,  along a given ray, we are solving a differential equation in a single variable $(z)$ whereas in the comoving frame we are solving a partial differential equation in two variables $(z, \nu)$ (Section~\ref{Sec_comp_obs}). The difference in accuracy is particularly noticeable for O stars  -- their narrow photospheric features become broadened as they are propagated (in both space and frequency) to the outer boundary.
\end{enumerate}

\begin{figure*}
\epsfig{file=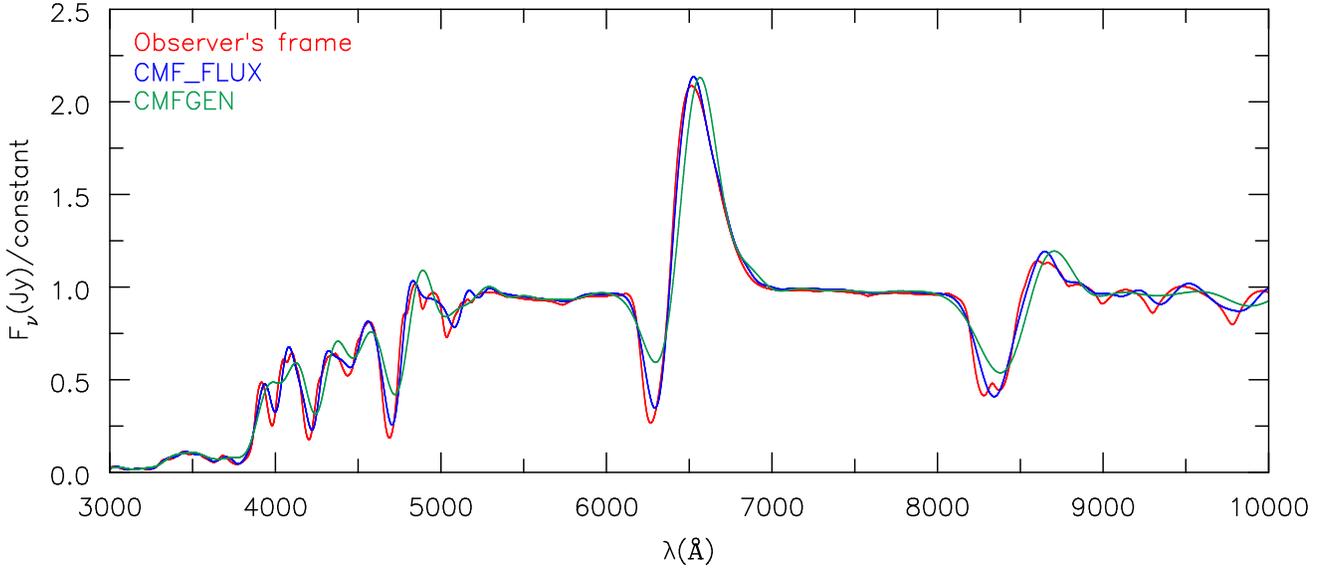,width=7.5cm, angle=-90}
\caption{Comparison of the predicted observed spectrum computed using an observer's frame calculation (red), {\sc cmf\_flux} (blue), and {\sc cmfgen}  (green) for a SN1987A like model at an age of 7.9\,days. The main characteristics of the spectra are in very good agreement but the observer's frame calculation is of higher accuracy, and suffers less smoothing than spectra computed by mapping comoving-frame intensities, computed with {\sc cmf\_flux} and {\sc cmfgen}, into the observer's frame.}
\label{fig_obs_comp_1}
\end{figure*}

\begin{figure*}
\epsfig{file=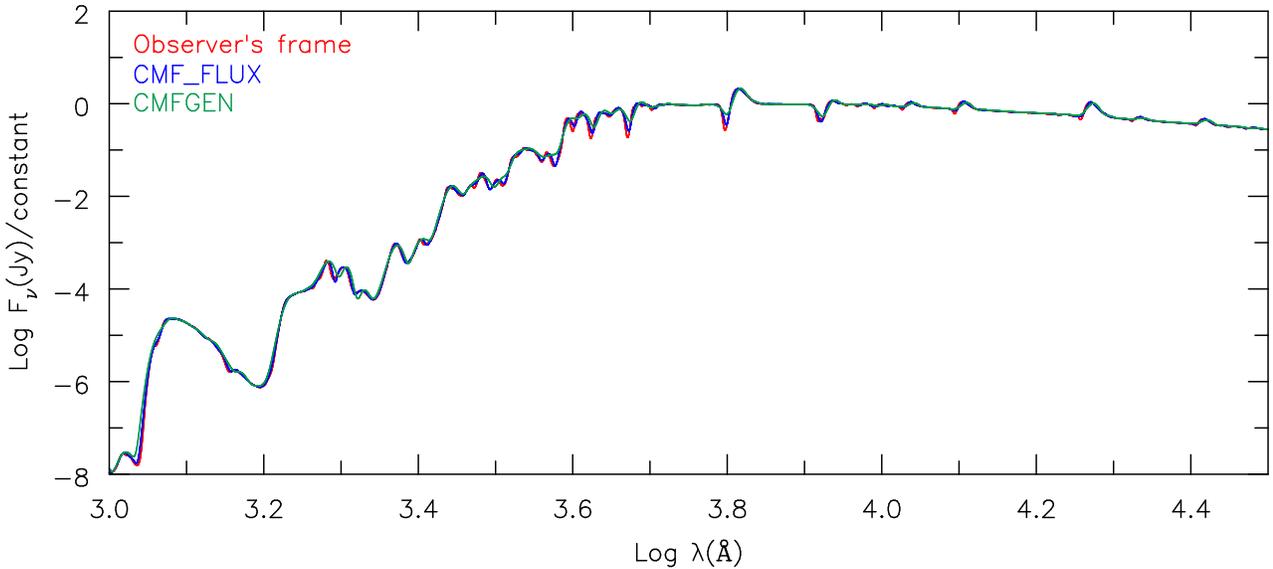,width=7.5cm, angle=-90}
\caption{As for Fig.~\ref{fig_obs_comp_1} but showing the logarithm of the flux over a broader wavelength range.}
\label{fig_obs_comp_2}
\end{figure*}

Ideally these techniques should give identical answers but in practice differences arise because the different techniques have different 
accuracies (as they use different numerical  techniques)  (\figs~\ref{fig_obs_comp_1} \&  \ref{fig_obs_comp_2}).
 

\begin{figure*}
\epsfig{file=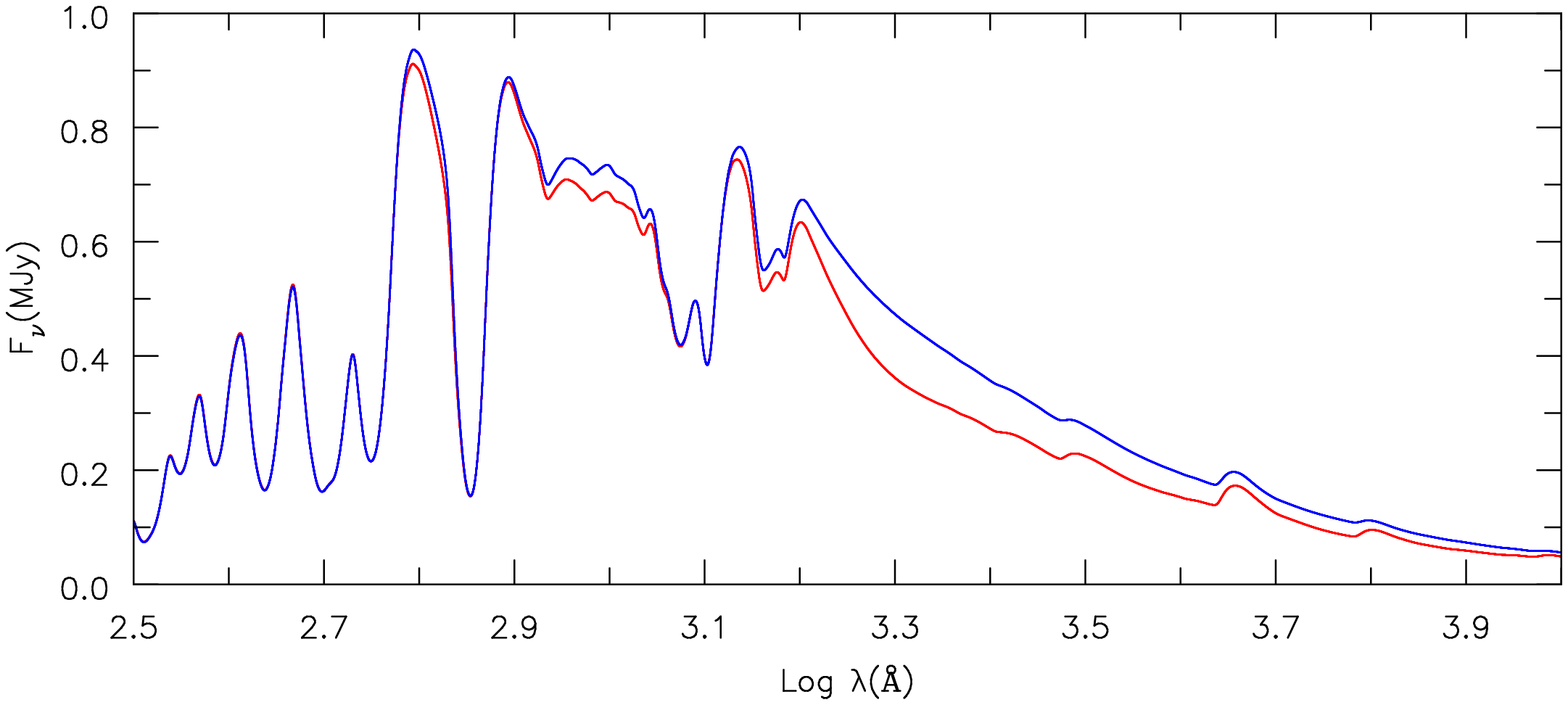,width=7.5cm, angle=-90}
\caption{Comparison of the predicted observed spectrum computed allowing
for time dependence in the moment equations (red) with the solution for the moments computed using the fully relativistic solution but ignoring time-dependence (blue).
Both spectra were computed in the observer's frame.}
\label{fig_obs_tdep}
\end{figure*}

At present we also ignore time dependence when computing observed spectra. This will be rectified in the future, but, for the same reasons outlined above (\S\ref{time_effects}), it is unlikely to have a large effect on observed spectra, except at the earliest times for H-rich core-collapse SNe when $J$ and $H$ are varying rapidly. For Type I SNe, the effects will be smaller since by one day the SN have already cooled, and their spectra are slowly evolving.  Note that time-dependence is taken into account in the calculation of the moments, and observer's frame spectra computed with these moments do differ from those computed where the time-dependence of the moments is ignored (Fig.~\ref{fig_obs_tdep}). The computation of the observer's frame spectrum is outlined in Section~\ref{Sec_comp_obs}.

\subsection{Comoving Frame computation of the observed spectrum}
\label{CMF_OS}

As we integrate from blue to red we store $I_o(t, \Rmax, \mu_o,\nu_o)$ at the outer boundary of the model. After the completion of each comoving frequency,  we check whether 
we can compute $I_s(t,\Rmax, \mu_s,\nu_s)$ for all $\mu_s$ for the next observer's frame frequency. If so, we compute $I_s(\Rmax, \mu,\nu)$ from $I_o(t, \Rmax, \mu_o,\nu_o)$ using the transformation from comoving to observer's frame (\eq~\ref{eqn_I_trans}) and linear interpolation in frequency. Because of the way we constructed the ray grid there is a one-to-one correspondence between $\mu_o$ and $\mu_s$, and hence no interpolation in $\mu$ is necessary. The observed flux, $F_s(t,\nu_s)$, at distance d is then computed using

\begin{eqnarray}
F_s(t,\nu_s) = {2 \pi \over d^2} \int_0^{p_{\rm max}} p\, I_s(t-d/c, R_{\rm max}, p,\nu_s) dp \,\,.
\label{eq_obs_flux}
\end{eqnarray}

\noindent
This integration is performed using numerical quadrature with $\mu_s$ as the integration variable
(note that $pdp =R^2_{\scriptsize \rm max}\, \mu_s d\mu_s$).

\subsection{Observer's frame calculation of the observed spectrum}
\label{Sec_comp_obs}

To compute observed spectra we use a modified form of {\sc cmf\_flux} which has been described by \cite{BH05_2D}. The
calculation of the observed spectrum proceeds as follows:

\begin{enumerate}

\item
 We solve the transfer equation in the comoving-frame using the assumption of coherent electron scattering in the comoving frame. This allows us to compute the mean intensity in the comoving frame. Along each ray we insert extra grid points -- typically we have a grid point every 0.5 Doppler velocities. The finer frequency grid adopted in {\sc cmf\_flux}  generally has only a minor influence on the computed spectra for Type II SNe.

\item
Using the technique of \cite{RH94_es} we compute the full, non-coherent and frequency dependent, electron-scattering source function.

\item
We resolve the transfer equation in the comoving frame, but this time we assume incoherent electron scattering, and hence use the electron-scattering source function previously computed.

\item
We repeat steps (ii) \& (iii) until convergence is obtained. This generally only requires a few iterations although in modeling the Type IIn SN 1994W a much larger number of iterations was needed. This was necessary because of the much higher line optical depths, at a given electron density, induced by the lower SN velocities \citep{DHG09_SN1994W}. In most SNe, frequency redistribution to the red, associated with the bulk motion of the gas, dominates and the non-coherence of electron scattering due to thermal motions is subdominant. In SNe where the spectrum formation region expands slowly, this non-coherent scattering is key and requires careful computation to yield accurate line-profile shapes.
\item
We map the comoving frame emissivities and opacities into the observer's frame using the following relations
\citep[see, e.g.,][]{Mih78_book}:

\begin{eqnarray}
\beta &=& v/c \\
\gamma &=& 1 / \sqrt{1 - \beta^2} \\
\chi_s(r,\mu_s,\nu_s) &=&(\nu_o/ \nu_s)\, \chi_o(r,\nu_o) \\
\eta_s(r,\mu_s,\nu_s) &=& (\nu_s/ \nu_o)^2 \, \eta_o(r,\nu_o) \\
I_s(r,\mu_s,\nu_s) &=& (\nu_s/ \nu_o)^3 I_o(r,\mu_o,\nu_o)  \label{eqn_I_trans}\\
\nu_s &=& \nu_o \gamma(1 + \mu_o \beta) \\
\nu_o &=& \nu_s \gamma(1 - \mu_s  \beta)
\end{eqnarray}

\noindent
In the above we use ``$o$'' to denote the comoving frame and ``$s$'' to denote the observer's (static/rest) frame. This mapping provides the emissivities and opacities necessary to solve the transfer equation in the observer's frame using the usual $(p,z)$ coordinate system. Along each ray (i.e., for each impact parameter $p$) we insert extra grid points so that the rapid variation of opacity and emissivity is adequately sampled -- typically we have a grid point every 0.25 Doppler velocities.

\item
We compute the observed flux using \eq~\ref{eq_obs_flux}.

\end{enumerate}


\section{Gray Transfer}
\label{Sec_gray}

To improve the speed of convergence it is desirable to have a good estimate of the temperature at each time step. In the inner, optically-thick regions, this is provided by solving the time-dependent gray transfer problem. The solution of the time-dependent gray transfer also provides a check on the solution of the full-moment equations.

Integrating \eqs~\ref{eq_zero_mom} and \ref{eq_first_mom} over frequency, and regrouping terms, we obtain
\begin{eqnarray}
  {1 \over cr^4}  {D(r^4 J)  \over Dt} + {1 \over r^2} {\partial (r^2 H)  \over \partial r}
  & = & \int_0^\infty \eta_\nu - \chi_\nu J_\nu d\nu \nonumber \\
   & = & \int_0^\infty \chi_\nu( S_\nu -  J_\nu) d\nu \nonumber \\
    & \simeq  &  \chip \int_0^\infty( S_\nu -  J_\nu) d\nu \nonumber \\
& \simeq  &  \chip( S- J) 
\label{eq_g0_mom}
\end{eqnarray}

\noindent
and
\begin{eqnarray}
  {1 \over cr^4}  {D(r^4 H)  \over Dt} + {1 \over r^2} { \partial(r^2 K)  \over \partial r}
  + {K - J \over r}  &=& - \int_0^\infty \chi_\nu H_\nu  \nonumber \\
  & \simeq & - \chir \int_0^\infty H_\nu  \nonumber \\
  & \simeq & - \chir  H   \,\,.
\label{eq_g1_mom}
 \end{eqnarray}

\noindent
where \chir\ is the Rosseland-mean opacity, and \chip\ is the Planck-mean opacity. Assuming LTE holds at depth we have $S_\nu=B_\nu$. Using
equation~\ref{eq_energy} with \ref{eq_g0_mom} we obtain



\begin{equation}
  {1 \over cr^4}  {D(r^4 J)  \over Dt} + {1 \over r^2} {\partial (r^2 H)  \over \partial r}
   =  {1 \over 4 \pi} \left(\dot{e}_{\rm decay} - \rho {De \over Dt} + {P \over \rho} {D\rho \over Dt}   \right)
 \label{eq_g3_mom}
 \end{equation}

\noindent
which follows from \eq~\ref{eq_energy}. After solving for $J$ we need to relate $J$ to $B$. Using \eq~\ref{eq_energy}, and since
$S=B$, we have

\begin{eqnarray}
 \int \chi_\nu B_\nu d\nu &=& \int \chi_\nu J_\nu d\nu \nonumber   \\
 & &  + {1 \over 4 \pi} \left(\dot{e}_{\rm decay} - \rho {De \over Dt} + {P \over \rho} {D\rho \over Dt} \right)
 \end{eqnarray}

 \noindent
 and thus
\begin{eqnarray}
 B  = \sigma T^4 &=& J  \nonumber  \\
 & &  + {1 \over 4 \pi \chip } \left(\dot{e}_{\rm decay} - \rho {De \over Dt} + {P \over \rho} {D\rho \over Dt} \right)
 \label{eq_BJ_rel}
 \end{eqnarray}

\noindent
where we have assumed $J_\nu \approx B_\nu$.

The solution of \eq~\ref{eq_g3_mom} is complicated by the presence of  the $\rho De/Dt - P D\ln \rho/Dt$ ($\equiv W$) term which depends on the unknown temperature structure. We handle the term by iteration --- that is, we compute the temperature structure for a given $W$ by solving \eqs~\ref{eq_g3_mom} and \ref{eq_BJ_rel},  update $W$, and then resolve \eqs~\ref{eq_g3_mom} and \ref{eq_BJ_rel}. Because of stability issues in the outer regions, we found it necessary to limit the changes in  $W$ to achieve convergence. To determine the change in $e$ we assume that the departure coefficients are constant -- in practice this is equivalent to assuming LTE in the inner regions where the gray approximation is valid.

Equations \ref{eq_g1_mom},  \ref{eq_g3_mom}, and \ref{eq_BJ_rel} represent 3 equations in 4 unknowns ($J, H, K, T$).  As for the frequency-dependent transfer equations we eliminate K using the Eddington relation $K=fJ$ where $f$ is obtained by solving the time-independent transfer equation, and for simplicity we neglect the curvature of the characteristics.

\begin{figure}
\epsfig{file=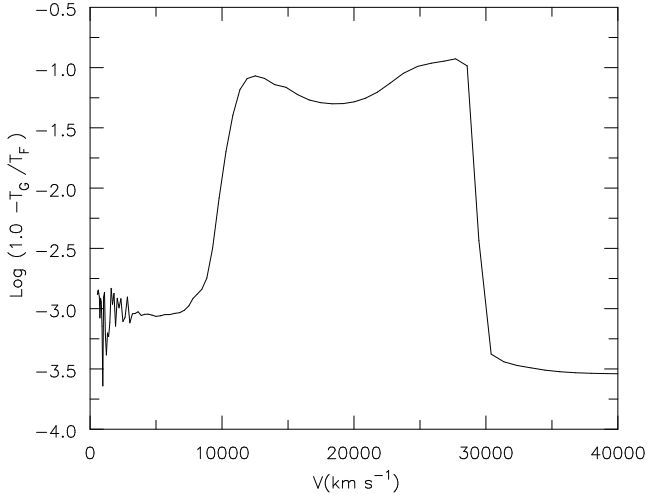,width=6.5cm, angle=-90}\caption{Comparison of the profiles for the  initial gray temperature $T_{\rm G}$ and the final model temperature $T_{\rm F}$. At depth the agreement is excellent.  The maximum difference between the two temperatures is about 10\%. At $v=$\,10\,000\,\kms\ the Rosseland optical depth is $\sim$\,12, while at 30\,000\,\kms\ it is $\sim$\,0.00035. Corrections for $v>$30\,000\,\kms\ are small because this model has an enforced minimum temperature.}
\label{fig_grey_comp}
\end{figure}

Since the Rosseland-mean opacity also depends on temperature, we perform several iterations of the gray-solution with updated Rosseland-mean opacities.  As the gray solution only provides a starting solution, high accuracy is unnecessary, and practical experience shows that the above procedure provides a very acceptable starting solution. As the gray solution does not hold everywhere, we only apply the gray-temperature structure to regions with an optical depth above $\tau_{\rm min}$. Below $\tau_{\rm min}$ we use $T$ from the previous time step. Around $\tau_{\rm min}$ we use a simple interpolation procedure so  as to switch smoothly between the two approximations. The choice of $\tau_{\rm min}$ is somewhat problematical. In O and WR stars we found that it was reasonable to adopt $\tau_{\rm min}\sim 1$ however  for Type II SNe we found that $\tau_{\rm min}\sim 5$ (or 10) provided better starting solutions. A comparison between the actual model temperature structure and the
gray temperature structure is show in Fig.~\ref{fig_grey_comp}.

\subsection{Global Energy Constraint}
\label{Sec_glob_en}

A global energy constraint can be obtained by integrating \eq~(\ref{eq_g3_mom}) over volume.  Multiplying \eq~\ref{eq_g3_mom} by $r^2$ and integrating  over $r$ we obtain

\begin{eqnarray}
 r_{\rm max}^2 H(r_{\rm max}) &=& r^2 H(r) \nonumber \\
&+& \int_{\hbox{$r$}}^{\hbox{$r_{\rm max}$}}
{r^2 \over 4 \pi  } \left(\dot{e}_{\rm decay} - \rho {De \over Dt} + {P \over \rho} \frac{D\rho}{Dt} \right) \nonumber \\
&-&   {1 \over cr^2}  {D(r^4 J)  \over Dt}\,\,  dr
\label{eq_global_en}
\end{eqnarray}

\noindent
In a static atmosphere, with $v=0$ this simply reduces  to $r^2H$=constant; and this is often used as the energy constraint in stellar atmosphere calculations at large $\tau$ ($\tau \gtrsim$\,1). This constraint has real physical meaning --- the flux at the outer boundary is set by the flux imposed at the inner boundary. However, for time-dependent SN calculations at early times, the constraint equation is, in some sense, merely a mathematical constraint.  Because of the finite speed of light coupled with the effects of diffusion, the outer flux is {\it independent} of the inner atmosphere at the current time step when the envelope has a ``large'' optical depth. Nevertheless, Eq.~\ref{eq_global_en} does allow a consistency check, and in complex codes consistency checks are always useful. In the outer envelope, and as the envelope thins, and in Type Ia SN with their much lower envelope masses, the constraint represented by Eq.~\ref{eq_global_en} becomes much more physically meaningful. Of course, depending on the balance between adiabatic cooling and other heating and cooling processes,
errors in the temperature of the inner envelope may manifest themselves at a later time.

In our original formulation, we integrated from $r_{\rm min}$ to $r$. However, we found such a formulation was not so useful since numerical errors (at large optical depths) tend to limit its usefulness. These numerical errors arise from the discretization ($J_\nu$ is defined on the grid while $H_\nu$  is defined at the midpoints), from the evaluation of the integrals, from the need for very high convergence, and from the technique used to difference the moment equations. For example, when we difference the zeroth moment equation we effectively replace
\begin{equation}
 \int_0^\infty \chi_\nu( S_\nu -  J_\nu) d\nu \nonumber
\end{equation}
 at each depth $d$  by
 \begin{equation}
  \int_0^\infty <\chi_\nu> ( S_\nu -  J_\nu) d\nu \nonumber
  \end{equation}
where $<\chi_\nu>$ is now averaged over several depths\footnote{The averaging in $\chi$ occurs because of the definition of the optical depth increments.}. By integrating inwards, we can more directly compare the constancy of the right-hand-side of equation  \ref{eq_global_en} with the quantity of interest --- the flux (or luminosity) at the outer boundary. 

\begin{figure*}
\includegraphics[scale=0.44, angle=-90]{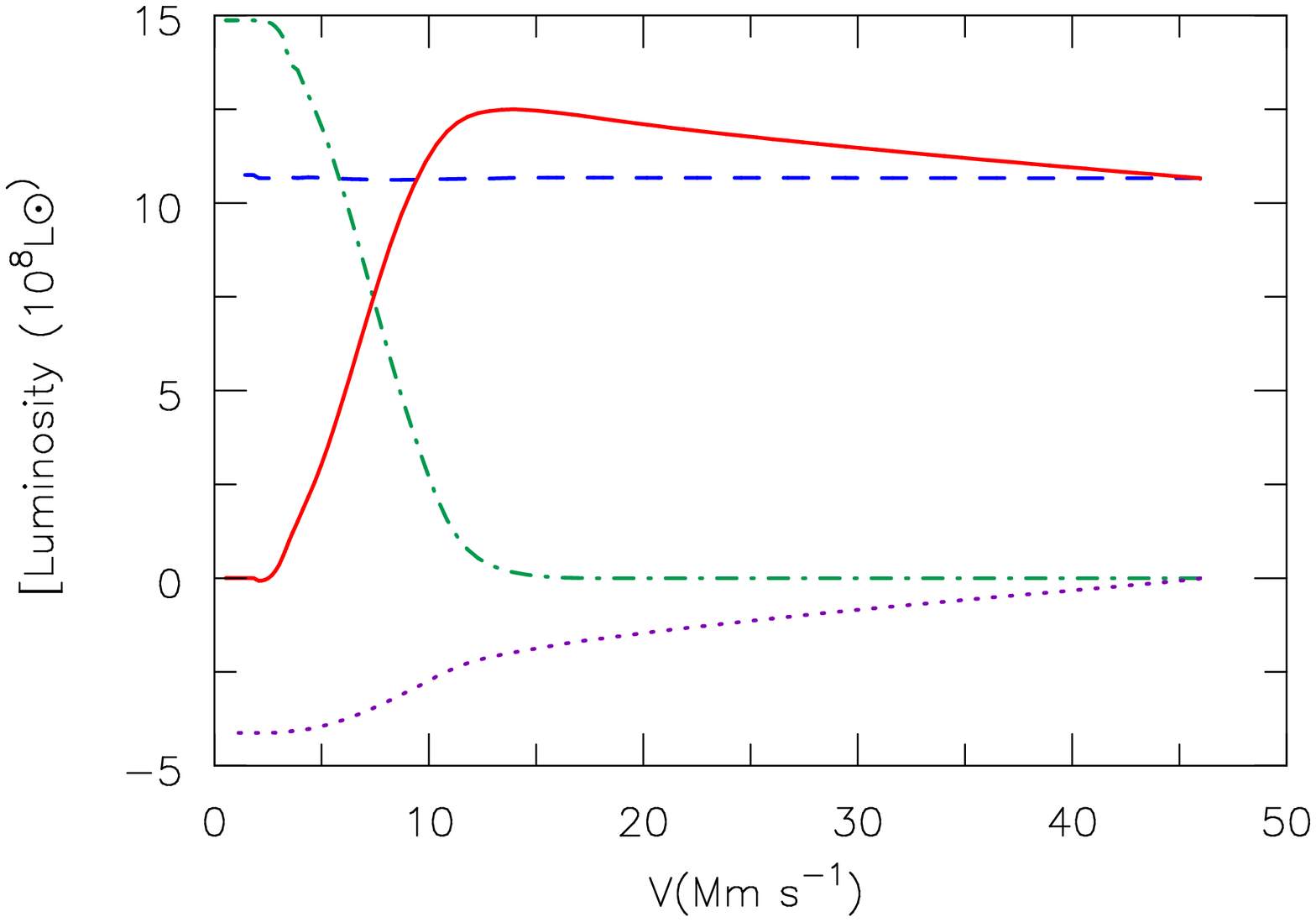}
\hfill
\includegraphics[scale=0.44, angle=-90]{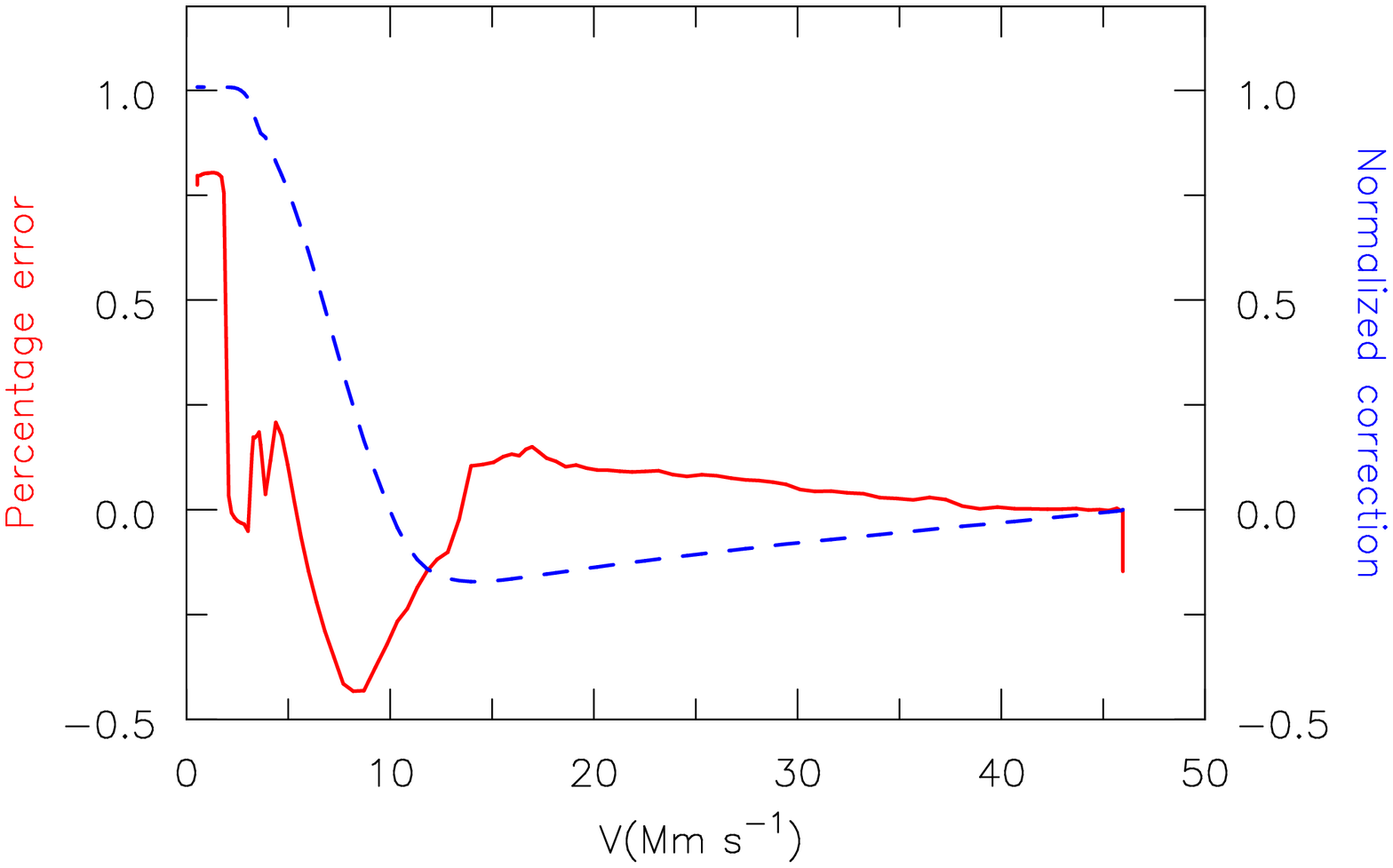}
\caption[]{IIllustration of the global energy constraint for a Type Ia SN model with local energy
deposition at an age of 60 days. In the left plot we see the depth-dependent luminosity (red; solid), the ``conserved luminosity" (blue; $---$), and the contributions to this
conserved quantity by energy from radioactive decay (green; $.-.-.-$), from the $Dr^4J/Dt$ term (purple; $...$). The gas term is negligible and is not shown. In the right plot we
see the percentage error in the conserved luminosity (red; solid), and the total correction to the luminosity 
as a function of depth, normalized by the outer boundary comoving-frame luminosity (blue; $---$). The model was computed using 114 depth points. }
\label{fig_Ia_cons}
\end{figure*}

\begin{figure*}
\includegraphics[scale=0.44, angle=-90]{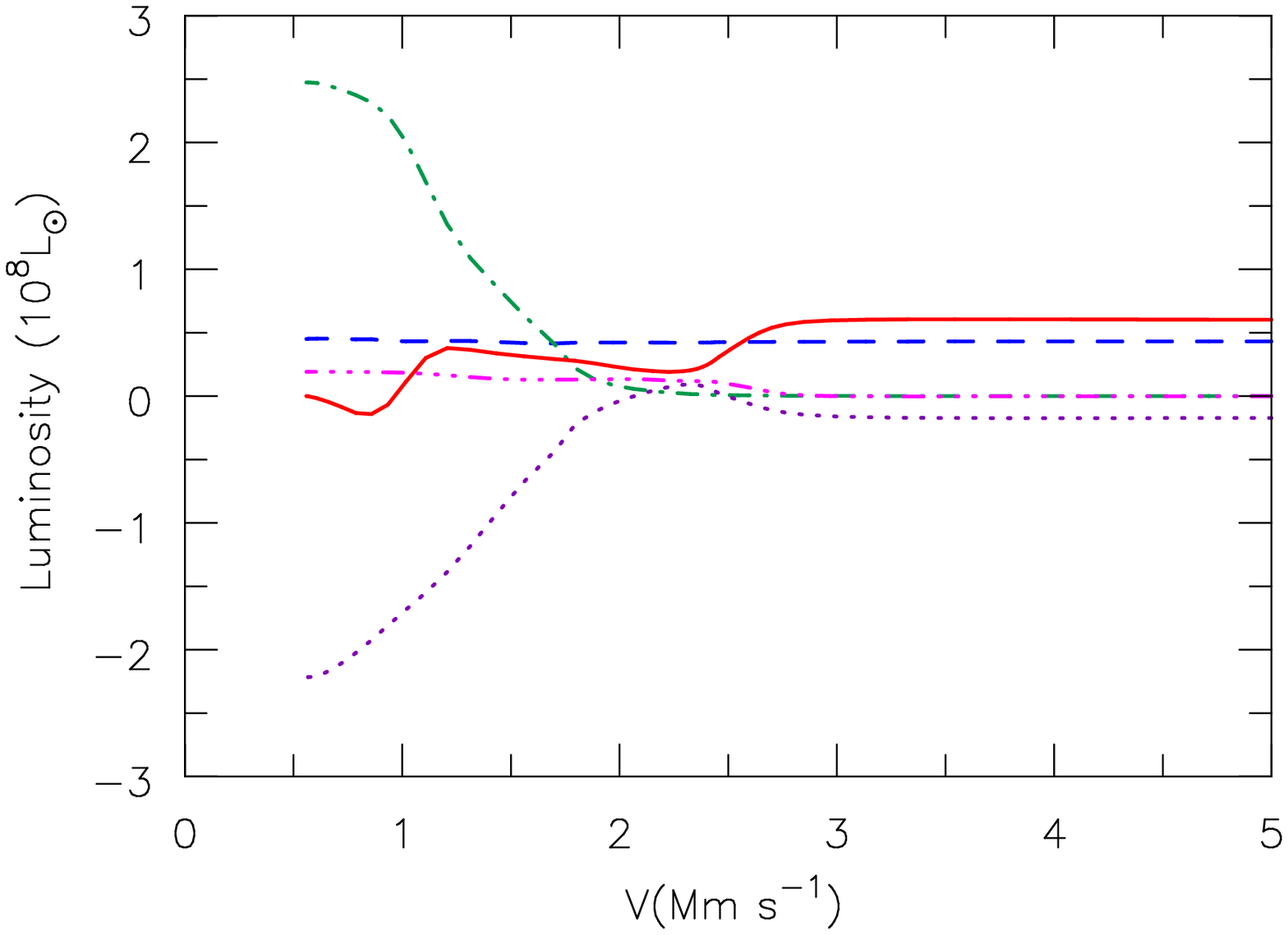}
\hfill
\includegraphics[scale=0.44, angle=-90]{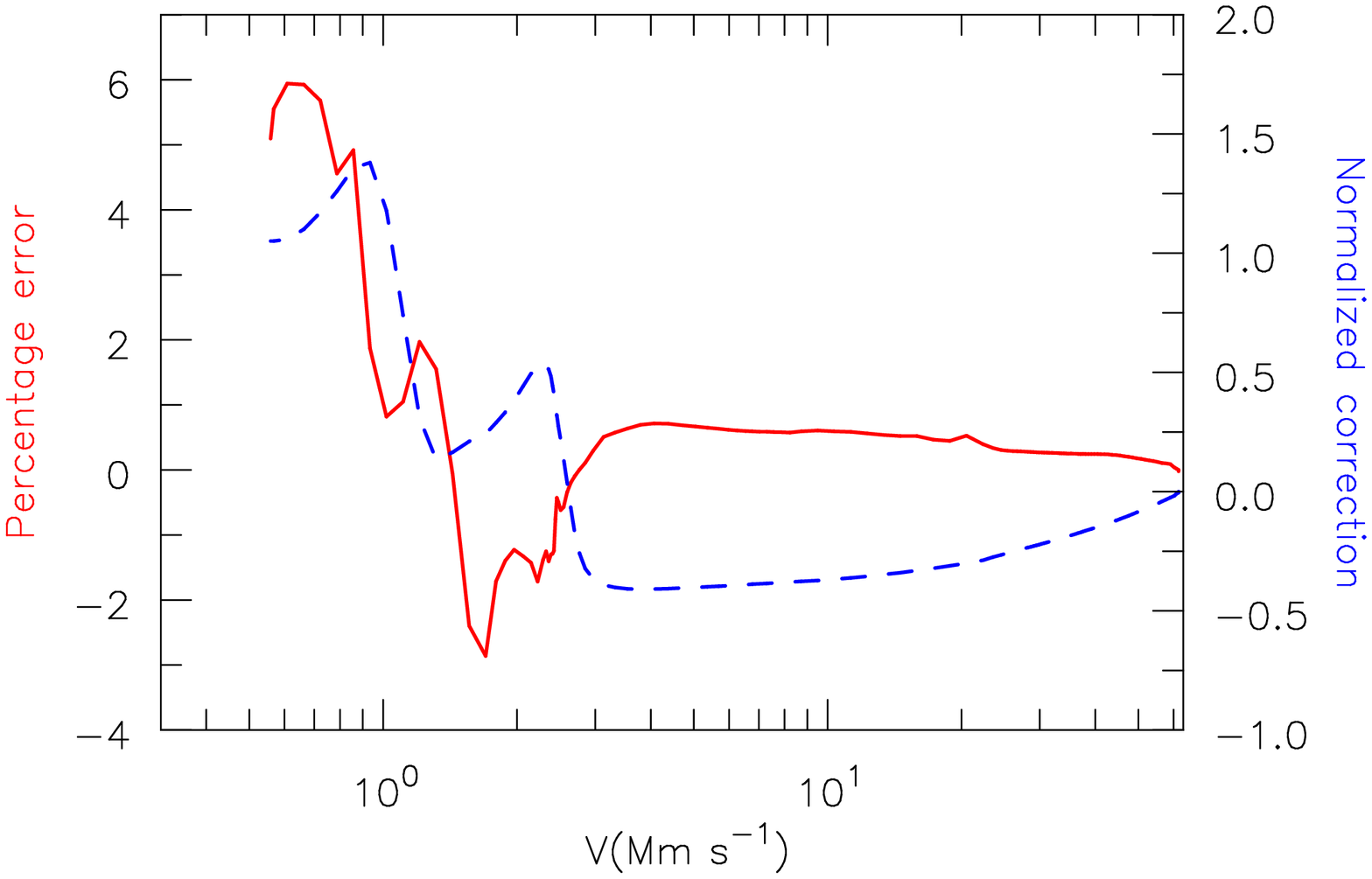}
\caption[]{IIllustration of the global energy constraint for a SN 1987A-like model with local energy
deposition at an age of 37 days. In the left plot we see the depth-dependent luminosity (red; solid), the ``conserved luminosity" (blue; $---$), and the contributions to this conserved quantity by energy from radioactive decay (green; $.-.-.-$),  $Dr^4J/Dt$ term (purple; $...$), and the gas term (pink;$-...$). In the right plot we see the percentage error in the conserved luminosity (red; solid), and the total correction to the luminosity as a function of depth, normalized by the outer boundary comoving-frame luminosity (blue; $---$). The model was computed using 126 depth points.}
\label{fig_II_cons}
\end{figure*}

In Fig.~\ref{fig_Ia_cons} and Fig.~\ref{fig_II_cons} we illustrate the global energy constraint as a function of depth for a Type Ia SN model, and model for SN 1987A. In the outer regions, constancy of the RHS is excellent, although small departures in the inner regions can be seen. These departures often arise from sharp changes in the material properties -- e.g., the H ionization front in Type II SN, or from  sharp changes in composition. By improving the resolution across these features we improve the accuracy of the global energy constraint. Through extensive testing we found that, in general, improving the accuracy with which the global energy constraint was satisfied did not significantly change predicted spectra, especially at earlier epochs.

\section{Future Work}
\label{Sec_future}

We have described the procedure we use to undertake time-dependent modeling of homologous SN ejecta. With our work, we solve self-consistently  the time-dependent \rte\ and its 0th and 1st moments, the time-dependent energy balance equation, and the time-dependent kinetic rate equations.   Our technique has been successfully used to model the Type II-peculiar SN 1987A \citep{DH10_time}, Type II-Plateau SNe \citep{DH11_SNII}, and Type IIb/Ib/Ic SNe \citep{DHL11_Ibc}. We are also using it to model Type Ia SNe \citep{DH12_SNIa} and extend our modelling of Type Ib/Ic SNe  \citep{DHLW12_SNIbc}.

With recent advances in hydrodynamic simulations there is increasing evidence that SNe are not spherically symmetric or homogeneous. There is also observational evidence for non-symmetric ejecta and for mixing in SNe. Ultimately 3D radiative transfer models may be needed to accurately model SNe ejecta. However, for a variety or reasons, 1D models still have a very important role. As our own work has shown, 1D models can be used to identify important fundamental physics such as the importance of time dependence in the rate equations for modeling Type II SN \citep{DH08_time} or the prediction that He\,{\sc i} emission lines can be excited in early spectra of Ib SN without the need for non-thermal processes \citep{DHL11_Ibc}. They also can provide fundamental insights into the abundance and physical  structure of ejecta -- information which is missing for most SN ejecta. Discrepancies between 1D models and observations can also provide crucial insights into the need for missing physics, and for features which can only be explained by 3D effects. Since hydrodynamical modeling cannot yet accurately predict the 3D structure of ejecta, 3D modeling is necessarily rich in free-parameters which can limit crucial insights. Importantly, 1D models also provide a testbed against which more complex models can be developed.

The present models are restricted to the case of  SN ejecta in homologous expansion. This is of little consequence for
Type Ia SNe, but restricts modeling for Type II-peculiar objects like SN 1987A to $t\gtrsim$\,1\,d, 
and to even later times ($t\gtrsim$\,10\,d) for ``generic'' Type II SNe of the plateau type. 
In order to overcome this limitation, work on a full-relativistic time dependent solver is in progress.

Another issue which we are addressing is the influence of non-thermal ionization and excitation. Such
processes are believed to be important for explaining the appearance of He\,{\sc i} lines in Type Ib SNe \citep{L91_HeI,DHL11_Ibc,DHLW12_SNIbc}, 
and for the production of H\one\ lines in the nebular phase of Type II SNe, and in 1987A \citep{XM91_87A_energetic,KF98_1987A,KF98B_1987A} .

In our work we treat  the whole ejecta. For Type I SNe, with their relatively low ejecta masses, this is a reasonable approach. However, for Type II SNe,  with their large ejecta masses, our approach may be an overkill. The inner layers of these models can be adequately described using a time-dependent  LTE gray model. A superior approach might be to combine a gray model with {\sc cmfgen}  modeling for optical depths  $\tau<\tau_{\hbox{{\scriptsize \rm max}}}$ (where $\tau_{\hbox{{\scriptsize \rm max}}}$ may be of the order of 100),  and use the gray solution to provide a lower boundary condition for {\sc cmfgen}. This would reduce the number of depth points in the {\sc cmfgen} models, and potentially the computational effort (since the model computation time scales as the square of the number of grid points).

With these tools we will be able to address many important questions related to SNe, their progenitors, and their mechanisms of explosion. For example, do stellar-evolution and hydrodynamical models of SNe agree with observations? What progenitors give rise to what SN types?  Do SN IIb/Ib/Ic arise from binary-star evolution? Does the composition of the outer unburnt ejecta correspond to predictions of stellar evolutionary models and stellar-atmosphere calculations of their progenitor class? 
Do we really understand the homogeneity of Type Ia SNe and the origin of the width-luminosity relation? Is there some signature in the spectra  of Type Ia SNe that may distinguish their origin from a single- or double-degenerate scenario?

\section*{appendix: \gray\ Monte-Carlo transport for SN ejecta}
\label{sec_gam_MC}

   We have two possibilities for the treatment of the decay energy in {\sc cmfgen}. We may deposit the entire decay energy locally and neglect any \gray\ transport, as described in Section~\ref{Sec_rad_energy}. This approximation is very good for hundreds of days in Type II SNe because unstable isotopes are produced explosively within the slower-expanding higher-density innermost layers of the massive ejecta. This is partially supported by the nebular-phase observations of  SN 1987A , which show the luminosity fading expected for full \gray\ trapping for up to $\sim$300\,d after explosion \citep{arnett_etal_89}. On the other hand \grays\ were detected as early as day 200, indicating that some
 \grays\ were escaping at that epoch, and by inference, that some non-local deposition was also
 occurring at that epoch. This, and the subsequent evolution of the  \grays\ and X-rays from SN 1987A,
 provide strong constraints on the mixing that has occurred \citep[e.g.,][]{KSN89_87A_xrays,AF89_SN9187A,FA89_SN1987A,BR95_87A_MC}.

  However, in the lower mass ejecta that characterize Type I SNe, either of thermonuclear or core-collapse origin, non-local energy deposition and even \gray\ escape can occur on a much shorter time scale \citep{DHL11_Ibc}. In Type Ia SNe, the deflagration/detonation produces \isoni\ at much larger velocities than in core-collapse SNe, in layers that can be well above the ejecta base depending on the details of the explosion \citep{khokhlov_91}. This property leads to \gray\ escape as early as the peak of the light curve \citep{hoeflich_etal_92}. Hence, to model such SN ejecta, one really needs to solve for the transport of \grays\ produced in the decay of unstable isotopes.

  Since our Monte Carlo code is to provide an energy-deposition distribution for {\sc cmfgen} runs, it uses the same model atom (i.e., species and ions), the same spatial grid, and assumes homologous expansion. The code treats only two-step decay chains but as many as desired, as in {\sc cmfgen}. The main focus for now though is on the decay chain
  associated with \isoni. The decay lifetimes,  \gray\ line energies (and probabilities), and positron energies are all taken from http://www.nndc.bnl.gov/chart/,
  themselves based on the work of \citet{HHZ87_nuc}. For \isoni\ and \isoco\  decays, these are in close agreement with the values 
  given by \citet{ambwani_sutherland_88,Nad94_nuc}. 
 The total decay energy for each channel, obtained by summing over all \gray\ lines and including positron energy, is used in {\sc cmfgen} when assuming
  local energy deposition. For non-local energy deposition, the Monte Carlo code is used instead to treat all individual contributions. 
  This ensures energy consistency whether we assume local or non-local deposition. 

  At present, we do not use the Monte Carlo procedure to select \gray\ lines but instead treat all important \gray\ photons associated with a given decay. 
 This is more time consuming but produces better statistics for weaker and/or improbable \gray\ lines.
 In practice, to reduce the burden on the number of \gray\ lines we follow, we treat only \gray\ lines  with a probability higher than 1\%. 
 This corresponds to 6 \gray\ lines for the decay of an \isoni\ nucleus and to 15 for \isoco. For \isoco\ decay, the missing  \gray\ lines cause a $\lesssim$3\% deficit
 in the decay energy, which we compensate by scaling all \isoco\ \gray\ line energies accordingly. 

  The Monte Carlo procedure we follow is analogous to that described by \citet{ambwani_sutherland_88} while the numerical approach closely follows
  \citet{hillier_91}. Although our SN models assume spherical symmetry, our calculations are performed on a 3D spatial grid.
  In the Monte Carlo technique, one has first to build probability functions
  for all important processes. For a given event described by the function $f$ of a variable $x$ over the interval [$x_{\rm min},x_{\rm max}$], we can generate $x$ from 
  \begin{equation}
  q =  \int_{x_{\rm min}}^{x} f(x) dx / \int_{x_{\rm min}}^{x_{\rm max}} f(x) dx
 \label{Eq_r_sel}
 \end{equation}
  where $q$ is a random number uniformly distributed between 0 and 1 \citep{hillier_91}.
  One can also bias the sampling of $f(x)$ by a user-specified function $b(x)$ so that $x$ is drawn from
 \begin{equation}
  q =  \int_{x_{\rm min}}^{x} b(x)f(x) dx / \int_{x_{\rm min}}^{x_{\rm max}} b(x)f(x) dx
  \end{equation}
  and that decay is then given a weight $1/b(x)$.

  We thus build probability distributions to describe the total number of decays, the relative number associated with each 2-step decay chain considered, 
  the relative number associated with each nucleus in each chain, and finally the radial distributions of such decays.

  For the simulation of \gray\ spectra, it is advantageous to bias the sampling of decaying nuclei to favor emission from the outer ejecta regions
  and thus enhance the statistics (reduce the noise) of the emergent \gray\ spectrum. This can help at early times when the bulk of unstable
  isotopes are located at large optical depths so that a very small fraction of \grays\ manage to escape. When biasing \gray\ emission for escape, we
  use the bias function $b(r)=\exp(-\tau(r))$ where $\tau(r)=\int_r^{\infty} \kappa(r') \rho(r') dr' $. For simplicity, we use a fixed mass-absorption
  coefficient $\kappa=$\,0.03\,cm$^2$\,g$^{-1}$ (\citealt{KF92_gam_rays}; this value does not need to be accurate).
  The associated energies for this decay at $r$ are then weighted by $1/b(r)$.
  In practice, this option is rarely used. Indeed, we are interested by the energy fraction that is deposited, not that which escapes.
  Furthermore, the eventual detection of \grays\ from SNe will most likely occur for an extragalactic event when the bulk of the released 
  energy escapes as the ejecta becomes transparent to \grays\ --- biasing emission is no longer needed at such times.
  
  Using a series of random numbers $q$, we select the decay chain, the nucleus that decays, and the
  decay location in spherical coordinates ($r$, $\theta$, $\phi$). $r$ is selected according to Eqn.~\ref{Eq_r_sel} with $x=r$ and $f(x)=4\pi r^2 \eta$ where $\eta$ is the emissivity. The angles are selected 
 using
  $$
  \cos\theta = 2q -1 \, , \,\,{\rm and}\,\,\, \phi = \pi (2q-1) \,,
  $$
  and a similar procedure is used to determine the original orientation of \gray\ emission.
 
  We then proceed to transport all \gray\ photons associated with that decay (note that we neglect the expansion of the ejecta over the life of such \grays\ --- the expanding ejecta appears as frozen to those \grays). If the decay also produces positrons, these are deposited locally as heat.
The location of the next scattering/absorption is conditioned by the opacity. We consider the two processes of Compton scattering
and photoelectric absorption (we neglect the opacity associated with pair production) and adopt the analytical formulae of \citet{KTN06_SN_MC}. 
Given an orientation 
we compute the associated total optical depth along that ray and direction out to the outer ejecta boundary.  We assume a hollow core for rays that cross the inner-boundary of the ejecta. Using another random number
$q$, the location of the next scattering/absorption is at an optical depth $\tau$ given by $q=1-\exp(-\tau) $ or $\tau=-\log(q-1)$.
In calculating these ray optical depths, we account for Doppler effects (redshifts) associated with expansion
and include both scattering and absorption opacities. If $\tau$ is greater than the maximum optical depth along the ray in the selected direction, the \gray\ photon escapes. Otherwise, we step along the ray to find the new location, change the photon energy to the scatterer's frame, and generate another random number to determine whether the photon is absorbed or scattered at that location. If the latter case is drawn, a new direction of scattering is sampled from the differential cross section for Compton scattering and the new energy of the photon is computed (the energy loss in the scattering is then deposited locally).

  In our calculations, we typically use 100,000 decays, which corresponds to about ten times as many \gray\ photons depending on 
the post-explosion time (i.e., whether it is primarily \isoni\ or \isoco\ nuclei that decay). The agreement between the local (computed analytically) and non-local (computed numerically with the Monte Carlo code) energy deposition profiles at very early times agree to within a fraction of a percent in most regions. The two can depart in regions where the original \isoni\ mass fraction is very low since too few decays will be triggered in those regions. Since we are mostly concerned with the bulk of the energy that is effectively deposited (and in appreciable amounts to influence the gas properties),  we are confident that the (originally) \isoni-rich regions are well sampled and the actual distribution profile is accurately determined.    

\begin{figure}
\epsfig{file=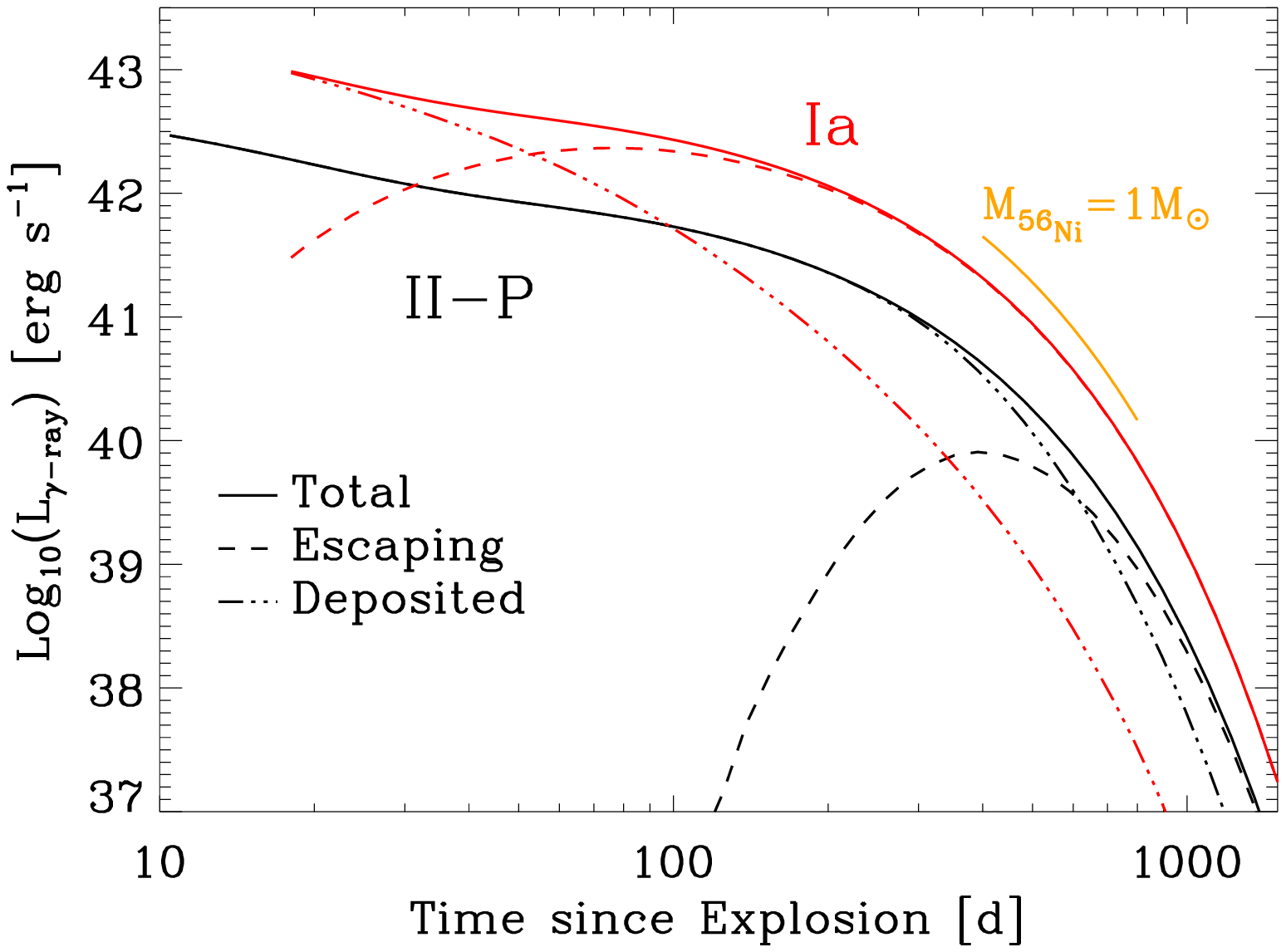,width=8.5cm}
\caption{Bolometric \gray\ light-curve from $\gtrsim$10\,d until 1500\,d after explosion for a SN Ia model (the model from \citet{KW07_Ia_width} with 0.49\,\Msun\ of \isoni\ initially) and a SN II-P model (the model s15e12 from \citet{DH11_SNII}, endowed with 0.087\,\Msun\ of \isoni\ initially). 
We show the evolution of the total decay energy (solid), the total decay energy deposited in the ejecta (dash-dotted), and the escaping \gray\ energy  (dashed). The orange line traces the radioactive-decay power of 1\,\Msun\, of \isoni\ produced initially in an explosion. 
\label{fig_lbol_gray}
}
\end{figure}

 We illustrate some interesting observable properties produced by such transport calculations. In \fig~\ref{fig_lbol_gray}, we illustrate
the evolution of the total, deposited, and escaping decay energy from ``representative''  SN Ia (the model with 0.49\,\Msun\ of \isoni\ of 
\citealt{KW07_Ia_width}) and II-P (the model s15e12 from \citealt{DH11_SNII}) models. In the Type II-P model, the larger mass, the slower
expansion and the deeper location of \isoni\ produced by the explosion lead to a much delayed emergence of \grays\ at a few hundred days after explosion. 
In contrast, the SN Ia model lets \grays\ start to escape as early as early as two weeks after explosion.
In each case, we illustrate the \gray\ spectrum one would observe at an epoch when \gray\ escape becomes strong. Because of the relatively
short half-life of \isoni\ nuclei, the \gray\ lines stem primarily from \isoco\ nuclei decay.

 We do not treat time-dependence in the Monte-Carlo code. This is a reasonable approximation since the light-travel time is small compared with the flow time, and because gamma-rays do not diffuse. As discussed by \cite{SSH95_gam_ray_dep} a \gray\ deposits half if its energy, on average, per scattering, and hence, after a few scatterings most of the gamma-ray energy has been deposited.

\begin{figure}
\epsfig{file=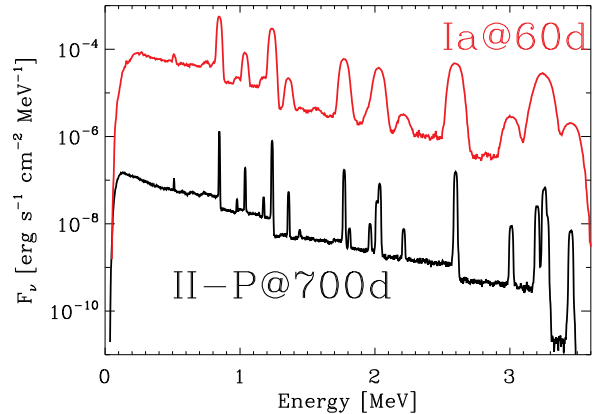,width=8.5cm}
\caption{\gray\ spectra for the SN II-P (black) and the SN Ia (red) models shown in Fig.~\ref{fig_lbol_gray} at post-explosion times of 60 and 700\,d. At such times, one sees primarily the \gray\ transitions associated with \isoco\ decay. For the calculations we assumed a SN distance of 10Mpc.
\label{fig_spec_gray}
}
\end{figure}

\section*{acknowledgements}
The authors would like to thank Douglas Miller for his contributions to the development of the formal solution with relativistic terms, and
the referee for their carful reading of the manuscrit and their thoughtful comments. DJH acknowledges support from STScI theory grant HST-AR-11756.01.A and NASA theory grant NNX10AC80G.  LD acknowledges financial support from the European Community through an International Re-integration Grant, under grant number PIRG04-GA-2008-239184. Calculations presented in this work were performed in part at the French National Super-computing Centre (CINES) on the Altix ICE JADE machine.

\bibliography{djdt_refs}

\label{lastpage}
\end{document}